\RequirePackage{fix-cm}
\documentclass[twocolumn,epjc3]{svjour3}  
\usepackage[T5,T1]{fontenc}

\smartqed  
\RequirePackage{graphicx}
\journalname{Eur. Phys. J. C}

\usepackage[utf8]{inputenc} 
\usepackage[T1]{fontenc}    
\usepackage{hyperref}       
\usepackage{url}            
\usepackage{booktabs}       
\usepackage{amsfonts}       
\usepackage{nicefrac}       
\usepackage{microtype}      
\usepackage{lipsum}
\usepackage{graphicx}
\usepackage{amsmath}
\usepackage{subcaption}
\usepackage[usenames,dvipsnames,table,modulo]{xcolor}
\usepackage[modulo]{lineno}
\usepackage{multicol}

\usepackage{widetext}
\usepackage{flushend}
\usepackage{cuted}

\usepackage{dblfloatfix}

\usepackage{tikz}
\usetikzlibrary{intersections}
\usetikzlibrary{decorations.pathmorphing}
\usetikzlibrary{shapes.geometric}
\usetikzlibrary{backgrounds}

\title{Low Energy Event Reconstruction in IceCube DeepCore}
\onecolumn

\author{R. Abbasi\thanksref{loyola}
\and M. Ackermann\thanksref{zeuthen}
\and J. Adams\thanksref{christchurch}
\and J. A. Aguilar\thanksref{brusselslibre}
\and M. Ahlers\thanksref{copenhagen}
\and M. Ahrens\thanksref{stockholmokc}
\and J.M. Alameddine\thanksref{dortmund}
\and A. A. Alves Jr.\thanksref{karlsruhe}
\and N. M. Amin\thanksref{bartol}
% ...
}
\authorrunning{IceCube Collaboration}
\thankstext{a}{also at Universit{\`a} di Padova, I-35131 Padova, Italy}
\thankstext{b}{also at National Research Nuclear University, Moscow Engineering Physics Institute (MEPhI), Moscow 115409, Russia}
\thankstext{c}{also at Earthquake Research Institute, University of Tokyo, Bunkyo, Tokyo 113-0032, Japan}
\institute{
% ...
Universit{\'e} Libre de Bruxelles, Science Faculty CP230, B-1050 Brussels, Belgium \label{brusselslibre}
% ...
\and Department of Physics, Loyola University Chicago, Chicago, IL 60660, USA \label{loyola}
\and Dept. of Physics and Astronomy, University of Canterbury, Private Bag 4800, Christchurch, New Zealand \label{christchurch}
% ...
\and Niels Bohr Institute, University of Copenhagen, DK-2100 Copenhagen, Denmark \label{copenhagen}
\and Dept. of Physics, TU Dortmund University, D-44221 Dortmund, Germany \label{dortmund}
% ...
\and Karlsruhe Institute of Technology, Institute for Astroparticle Physics, D-76021 Karlsruhe, Germany  \label{karlsruhe}
% ...
\and Bartol Research Institute and Dept. of Physics and Astronomy, University of Delaware, Newark, DE 19716, USA \label{bartol}
% ...
\and Oskar Klein Centre and Dept. of Physics, Stockholm University, SE-10691 Stockholm, Sweden \label{stockholmokc}
% ...
\and DESY, D-15738 Zeuthen, Germany \label{zeuthen}
}

\begin{document}
\maketitle
\twocolumn

\begin{abstract}

The reconstruction of event-level information, such as the direction or energy of a neutrino interacting in IceCube DeepCore, is a crucial ingredient to many physics analyses.
Algorithms to extract this high level information from the detector's raw data have been successfully developed and used for high energy events. In this work, we address unique challenges associated with the reconstruction of lower energy events in the range of a few to hundreds of GeV and present two separate, state-of-the-art algorithms. One algorithm focuses on the fast directional reconstruction of events based on unscattered light. The second algorithm is a likelihood-based multipurpose reconstruction offering superior resolutions, at the expense of larger computational cost.

\end{abstract}


\newcommand{\vertex}{\ensuremath{V_\textrm{xyzt}}}
\newcommand{\Ecscd}{\ensuremath{E_\textrm{cscd}}}
\newcommand{\Etrck}{\ensuremath{E_\textrm{trck}}}
\newcommand{\ltrck}{\ensuremath{\ell_\textrm{trck}}}
\newcommand{\nuecc}{$\nu_e^{\textrm{CC}}$ }
\newcommand{\numucc}{$\nu_\mu^{\textrm{CC}}$ }
\newcommand{\nutaucc}{$\nu_\tau^{\textrm{CC}}$ }

\section{Introduction}

\begin{figure}[h!]
    \centering
    \includegraphics[width=1\linewidth]{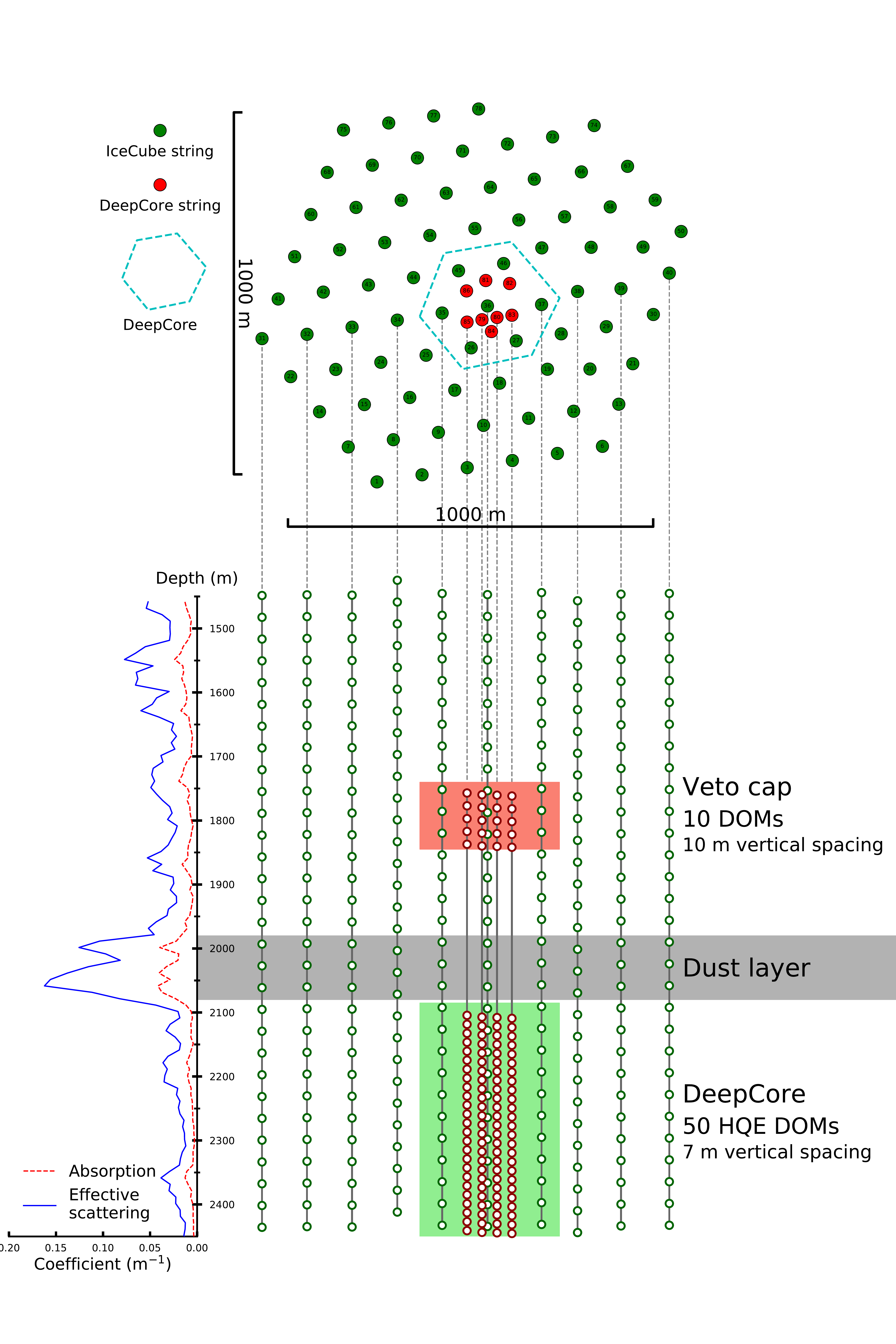}
    \caption{Schematic view of the IceCube in-ice detector array. The DeepCore sub-array, located at the bottom center, features denser module spacing and allows for the detection of lower energy events compared to the rest of IceCube. Also shown are the depth-dependent effective scattering and absorption coefficients  \cite{Chirkin:2013lpu}. The zone labeled as ``Dust layer'' is of inferior optical quality with increased scattering and absorption. The DeepCore array is located below this in the clearest part of the ice.}
    \label{fig:icecube}
\end{figure}

\begin{figure}[h]
\centering
\begin{subfigure}{\linewidth}
  \centering
  \includegraphics[width=.8\linewidth]{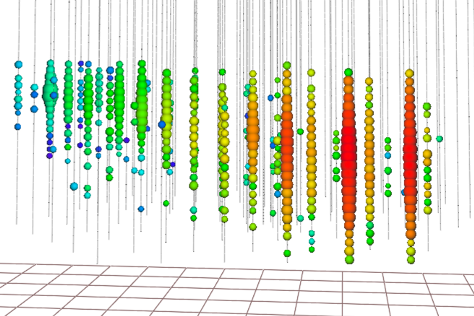}
  \caption{A high energy event (IceCube-170922A) with estimated energy of about 290 TeV. \cite{IceCube:2018dnn}}
  \label{fig:txs}
\end{subfigure} \\
\begin{subfigure}{\linewidth}
  \centering
  \includegraphics[width=.8\linewidth]{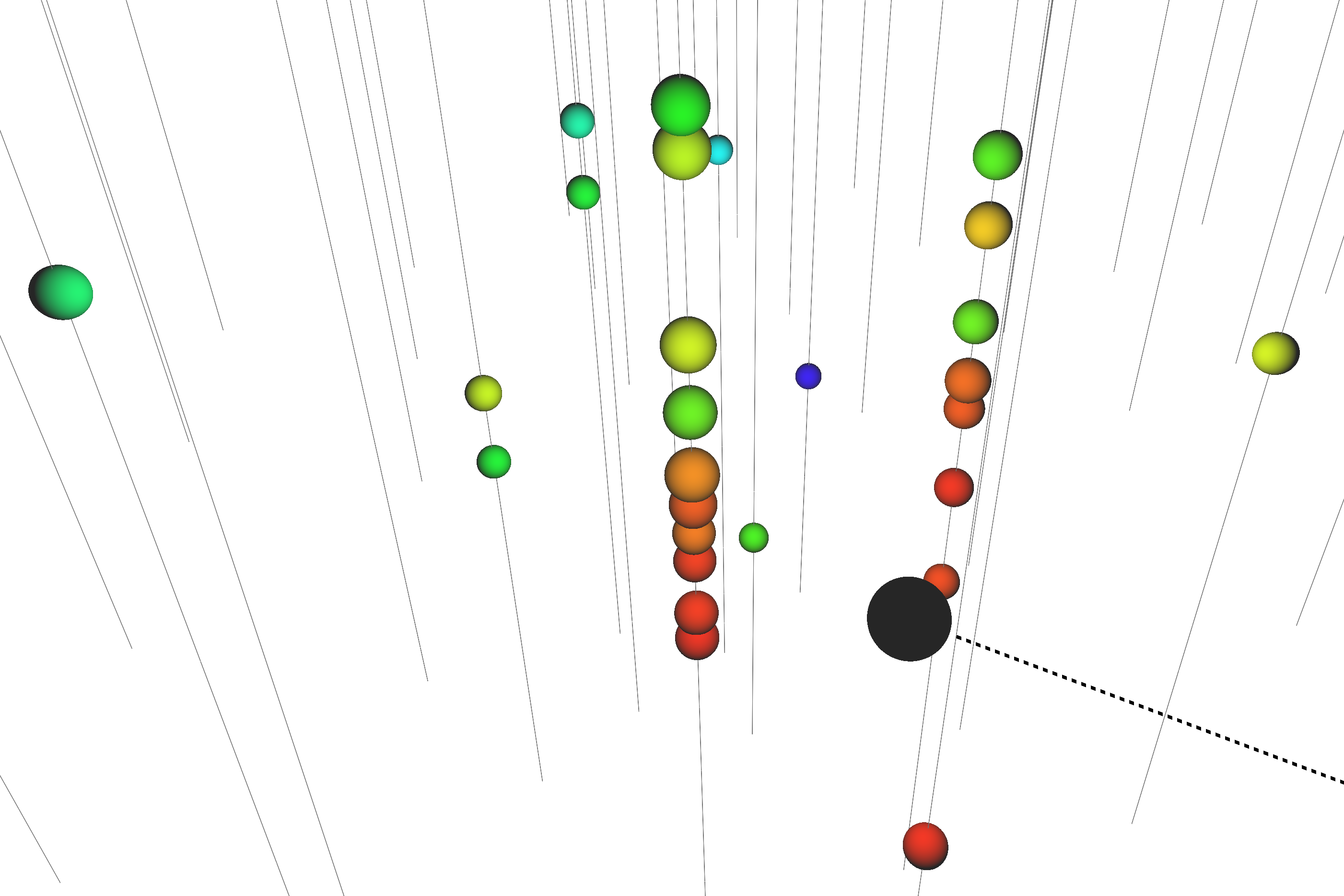}
  \caption{A simulated low energy \nuecc neutrino interaction with an energy of 25 GeV inside the DeepCore sub-volume. The true neutrino direction is shown as the black dashed line, and the vertex as the black, round marker. }
  \label{fig:cscd}
\end{subfigure} \\
\begin{subfigure}{\linewidth}
  \centering
  \includegraphics[width=.8\linewidth]{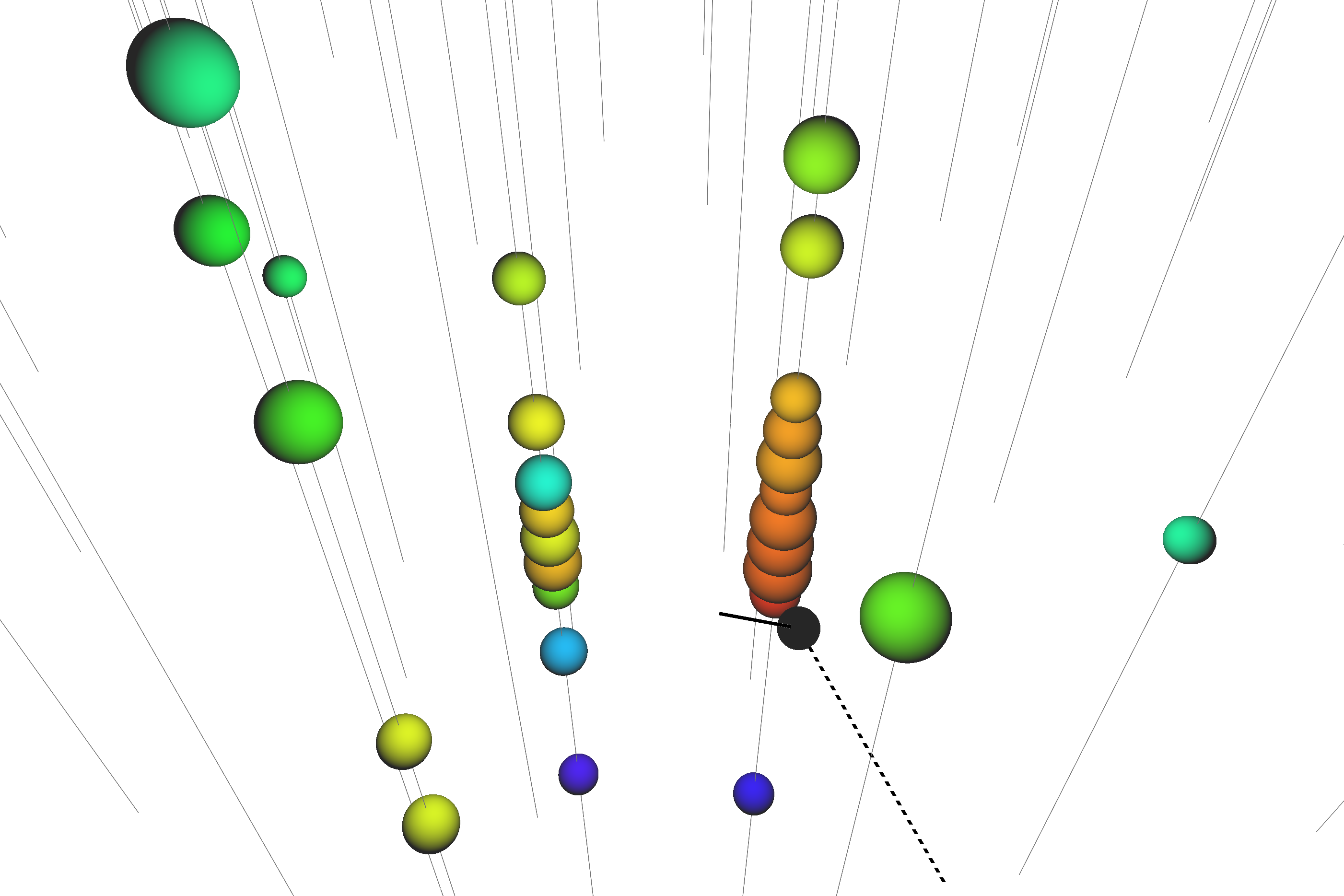}
  \caption{A simulated low energy \numucc neutrino interaction with an energy of 25 GeV inside the DeepCore sub-volume, with an outgoing 8 GeV $\mu^-$ indicated with the black, solid line (track).}
  \label{fig:trck}
\end{subfigure}
\caption{Example event displays: The size of spheres represent the amount of light observed in each DOM; larger spheres correspond to more light observed.
The color represents the timing of the hit, where the earliest hits are colored in red and the latest hits in blue. Low energy events have significantly fewer hits.}
\label{fig:event_display}
\end{figure}

The IceCube Neutrino Observatory \cite{Aartsen:2016nxy} at the South Pole registers Cherenkov light from particle interactions in the ice with its 5160 digital optical modules (DOMs). The modules are deployed at depths between 1450 and 2450 meters, covering a volume of a cubic-kilometer of ice, and arranged in 86 vertical strings (see Fig. \ref{fig:icecube}).
Eight dedicated strings with denser spacing and photomultiplier tubes (PMTs) with higher quantum efficiency form, together with the 7 central IceCube strings, the DeepCore sub-array \cite{DeepCore}.
The module density in DeepCore is about five times greater than the rest of IceCube, which allows for a much lower energy detection threshold of a few GeVs. Most of the recorded events in IceCube are from atmospheric muons or random triggers caused by dark noise, while only a fraction are from atmospheric neutrinos and a few events can be attributed to astrophysical neutrinos.
In all cases, the event information -- after some basic processing -- is represented as a series of \textit{pulses} per optical module, with each pulse consisting of a time and a charge.
Figure \ref{fig:event_display} shows event views for one high and two low energy events. 
In the process of event reconstruction, this raw pulse information is used to estimate properties of the interaction, such as interaction vertex, energy, and direction (angles).

Multiple algorithms to infer such quantities in IceCube (and its predecessor experiment, AMANDA~\cite{Andres:2001ty}) have been successfully developed and used, including those reported in Refs.~\cite{Ahrens:2003fg,Aartsen:2013vja,Abbasi_2013,Aartsen:2013bfa} and recently Refs.~\cite{Abbasi:2021ryj,IceCube:2021oqo}.
These algorithms, however, are geared towards high energy interactions for neutrinos in the TeV-PeV energy range such as the one shown in Fig.~\ref{fig:txs}.
For oscillation physics or dark matter searches, for instance, energies in the GeV range are of interest.
Such events create much less light in the detector resulting in sparse pulses (see Figs.~\ref{fig:cscd} \& \ref{fig:trck}), and hence involve different challenges for reconstruction than in the high energy regime. At the same time, a large number of events are collected at these low energies, with typical analysis samples easily exceeding a hundred thousand observed events, and simulated events used in analyses amounting up to $\mathcal{O}(10^8)$ events that require reconstruction.

In this article, we describe two distinct strategies addressing the challenges of extracting event-level reconstruction information from sparse data at high computational speeds:

\begin{itemize}
    \item Selecting events dominated by unscattered light results in a comparably small, but easy-to-reconstruct and robust set of events, and allows for fast turn-around physics analyses. The corresponding algorithm is discussed in Sec.~\ref{sec:santa}, representing an enhancement of an existing direction reconstruction \cite{Aguilar:2011zz}, and is limited to events passing a selection of direct light. This type of reconstruction was used, for instance, for the results of Ref.~\cite{Aartsen:2014yll}.

    \item To explore a larger, and more inclusive amount of detector data, the second algorithm, discussed in Sec.~\ref{sec:retro}, offers a multipurpose event reconstruction applicable to any event, and yields more comprehensive information at the expense of higher computational cost. A predecessor to the algorithm discussed is described in Ref.~\cite{Leuermann}. This type of reconstruction was used, for example, for the results of Refs.~\cite{Aartsen:2017nmd,Aartsen:2019tjl,IceCube:2019dyb,IceCube:2021abg}.
 
\end{itemize}

\subsection{Key Observables}
\label{sec:observables}
We start with discussing important observable properties for neutrino events in IceCube DeepCore. 
For many physics analyses, reconstructed event-level information, such as estimates for an event's deposited energy and point of interaction (vertex) in the ice, are crucial ingredients. Such information is often used in the event selection to remove background events (e.g. atmospheric muons or uncontained events), but is also used directly in an analysis itself.

As an example, we can illustrate this for oscillation physics using the two-flavor vacuum oscillation probability for a neutrino produced in flavor eigenstate $\alpha$ to be observed in the different eigenstate $\beta$:
\begin{equation}
    P_{\alpha\to\beta} = \sin^2(2\theta) \sin^2 \left(\frac{\Delta m^2 L}{4E}\right).
\end{equation}
This probability depends on the neutrino's energy, $E$, the length between production and interaction, $L$, the mixing angle, $\theta$, and mass  splitting, $\Delta m^2$  (all in natural units).
In order to resolve the oscillations and infer the parameters of interest, $\theta$ and $\Delta m^2$, precise estimates of the energy, $E$, the distance, $L$, and the flavors, $\alpha$ and $\beta$, are desired.
For atmospheric neutrino oscillations, our estimate of the energy of an incoming neutrino is the deposited energy of the event.
For the purposes discussed in this work, this ``deposited'' energy is defined as the difference in energy of the incoming and any outgoing neutrino. In the case of charged-current (CC) tau neutrino interactions, \nutaucc, half the energy of the the tau is subtracted to approximate the invisible energy of the daughter neutrino in the tau decay.
For atmospheric neutrinos, the propagation length can be calculated using the zenith direction, $\vartheta$, for which an approximate relation is $L \sim \cos{\vartheta}$.
For atmospheric neutrinos, the initial flavor, $\alpha$, at the production site is not known at the event level, and the expected fluxes of each flavor are modeled theoretically.
This means for the oscillation analysis example, estimates of $E$, $\vartheta$ and a handle on the interaction flavor $\beta$ (see Sec.~\ref{sec:signatures}) are desired.


\subsection{Event Signatures}
\label{sec:signatures}


Detectable neutrino interactions in IceCube can be grouped into two general categories: those which only contain a so-called ``cascade'' signature, and those with an additional ``track'' (see Figs.~\ref{fig:cscd}~\&~\ref{fig:trck}, respectively).
We categorize any particle showers as cascades. These include hadronic showers caused by deep inelastic scattering, and electromagnetic showers induced by the electron from \nuecc interactions.
Relativistic, charged particles ($\beta > 1/n$) in such showers emit Cherenkov radiation that can be detected by IceCube DOMs, and for energies between 1–100~GeV, such showers extend approximately 4–6~m in length
\cite{Li:2015kpa}.
In contrast to hadrons and electrons, muons have the potential to travel a sizable distance in the ice while radiating Cherenkov light, leading to a track signature in the detector.
In general \numucc interactions will produce muons, and about 17\% of tau leptons produced in \nutaucc interactions decay to muons, while other interactions result only in cascades \cite{PDG}.
We are interested in distinguishing events containing tracks from pure cascade events in order to get a handle on the flavor of the neutrino that we observe.

\subsection{Raw Data / Pulses}
\label{sec:pulses}

Photo electrons created in the PMTs in IceCube's DOMs generate a voltage at the PMT's anode that, when exceeding a discriminator threshold, triggers a digitization process of the signal via analog to digital converters (ADCs). This digitization process involves multiple gain channels, and different digitizers operating at different sampling rates. \cite{Abbasi:2008aa}.  The resulting waveforms then undergo a processing step known as {\it wavedeform} \cite{Aartsen:2013vja} that extracts the time and charge of individual single photo electrons (SPEs) \cite{IceCube:2020nwx} from the raw data, resulting in a collection of pulses, ${t_i}$, with charges ${q_i}$.
Before applying reconstruction algorithms, these pulse series typically undergo a cleaning algorithm that reduces isolated pulses that are created by dark noise.
A clustering approach requires all pulses to be causally connected within a distance of at most 150\,m and a time difference of at most 1\,$\mu$s to at least one other pulse in the cluster.
The resulting series of pulses per event serve as the input data to the reconstruction algorithms described next in this article.
For a typical oscillation event sample, as used in Sec.~\ref{sec:performance}, and after the cleaning described above is applied, the average events contain a total of 17 pulses, distributed over 14 DOMs and 6 strings.

\section{\textsc{santa} --- Avoiding Scattered Light}
\label{sec:santa}
The Single-string Antares-inspired Analysis (\textsc{santa})\cite{Garza2014Measurement} is an algorithm for the reconstruction of track directions. It is an improved and adapted version of a similar approach originally used at the ANTARES neutrino telescope~\cite{Aguilar:2011zz}, and developed with the intention of building a fast reconstruction
method capable of online monitoring.
A discussion of the working principles behind the algorithm and the improvements made with respect to the original implementation is the scope of this section. 

The reconstruction algorithm is preceded by a cleaning routine which removes pulses produced
by light that has undergone significant scattering as it traveled
through the ice, leaving only minimally scattered photons in the event.
This hit selection routine is described in Sec.~\ref{subsec:hit-selection}.
With the scattering removed, we calculate the expected observation time for the
Cherenkov photons geometrically as discussed in
Sec.~\ref{subsec:geometric-time-derivation}. Finally, we run a $\chi^2$-fit with respect to the observed and predicted observation time with an additional regularization term involving the observed charge as described in Sec.~\ref{sec:santa-loss}.

\subsection{Hit selection}

\label{subsec:hit-selection}The first step of the \textsc{santa} reconstruction
is the selection of minimally scattered photons from all of the observed
pulses by removing those pulses that are likely to have undergone a significant amount of scattering.

We combine the pulse series recorded by each activated DOM to a \emph{hit} with the time of the first pulse and the total charge of all pulses. All subsequent cleaning and reconstruction steps are applied to these combined hits. To remove scattered light, we make use of the fact that the largest possible delay between hits that are produced by \emph{unscattered} light on two different DOMs, $i$ and $j$, on the same string, is the time it takes for a directly up- or down-going light front to travel from one DOM to the other, $\tau_{ij}=|\Delta z_{ij}|/c_{\mathrm{ice}}$, where $|\Delta z_{ij}|$ is the distance between DOMs $i$ and $j$ and $c_{\mathrm{ice}}$ is the speed of light in ice. If the time delay between two hits, $\Delta t_{ij}$, is larger than the maximum delay, $\tau_{ij}$, then we know that the light must have undergone some amount of scattering. As a starting point for the hit selection, we choose the hit with the highest charge on each string, $i=0$, and first remove any earlier hit, $j$, where $-\Delta t_{0j} > \tau_{0j}$. From there, the algorithm iterates through every hit, $i$, and removes any other hit, $j$, where $\Delta t_{ij} > \tau_{ij}$. If fewer than 3 hits remain on a string, the entire string is removed from the event. If less than 5 hits remain in the event, it cannot be reconstructed. This is a simplified version of the cleaning procedure described in Ref.~\cite{Garza2014Measurement} and leaves more scattered light in the events. This is compensated for by the addition of the robust loss function (Sec.~\ref{sec:robust-losses}). In this configuration, we can reconstruct about 10\% more \numucc events than with the original implementation~from Ref.~\cite{Garza2014Measurement} at a similar resolution. In the example event fits in Figs.~\ref{fig:santa-single-string-example} \& \ref{fig:santa-multi-string-example}, the hits that are removed by the hit selection are crossed out.

\subsection{Single-String vs. Multi-String}

\begin{figure}[h]
    \centering
    \includegraphics[width=\linewidth]{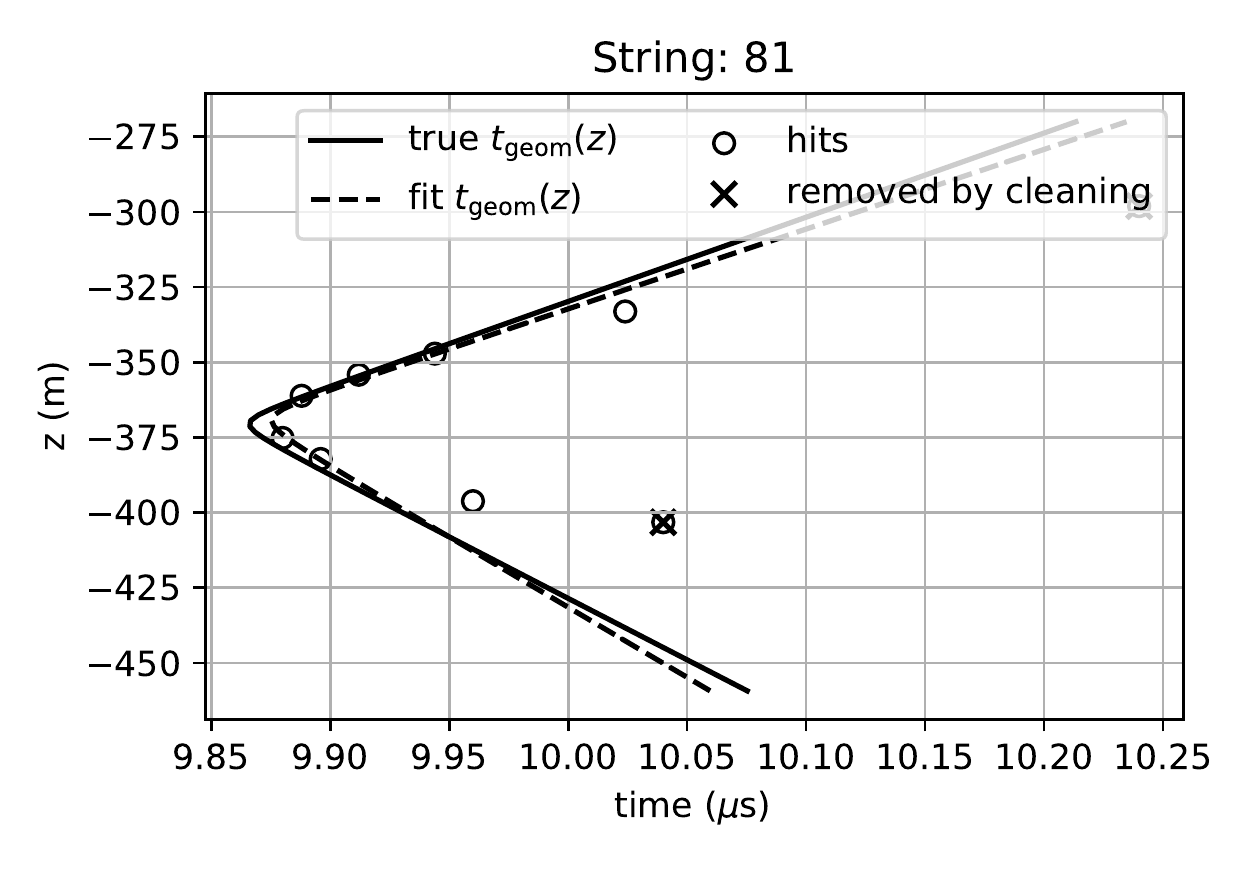}
    \caption{Example of a \numucc event reconstructed with \textsc{santa} on a single string. Circles show each hit, where the z-coordinate is the position of the DOM and the time is the time of the first observed pulse in that DOM.}
    \label{fig:santa-single-string-example}
\end{figure}

\begin{figure*}[ht]
    \centering
    \includegraphics[width=\textwidth]{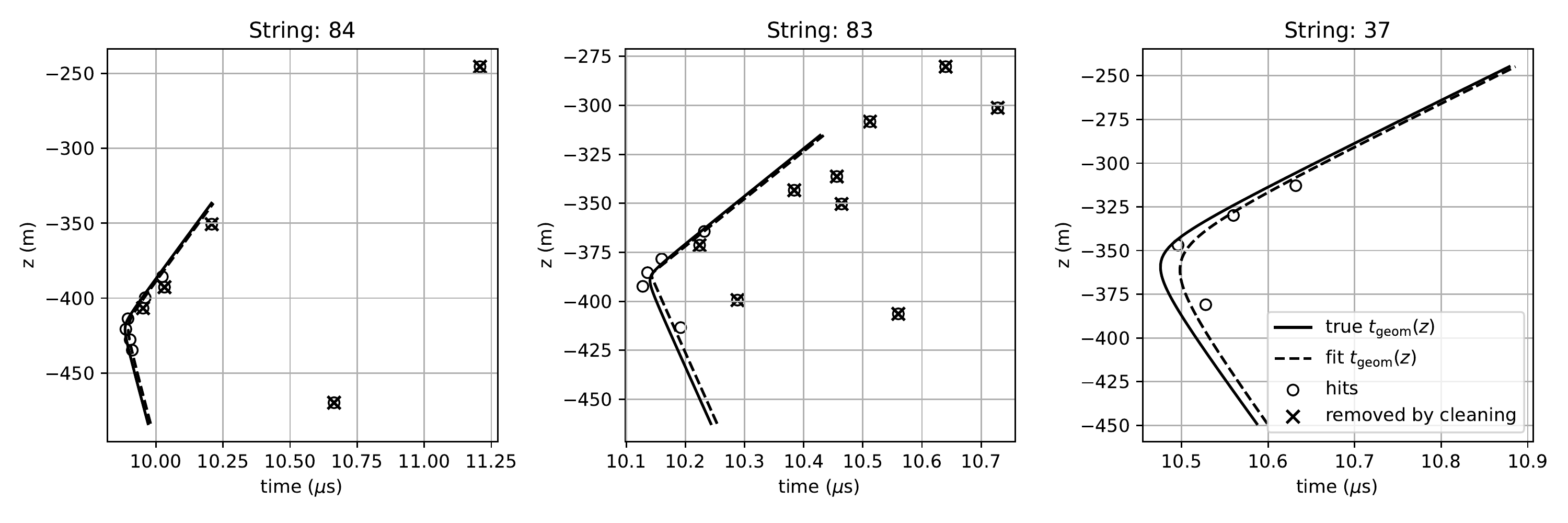}
    \caption{Example of a \numucc event reconstructed with \textsc{santa} with hits on several strings. Strings 84, 83 and 37 are spaced $\sim80\,\mathrm{m}$ apart from each other and form a highly obtuse triangle.}
    \label{fig:santa-multi-string-example}
\end{figure*}

After the hit-cleaning procedure, passing events fall
into two basic categories that are reconstructed differently. The first category is \emph{multi-string}
events that contain observed charges in modules on two or more strings
of the detector. Since a string is removed entirely from an event
if it has less than three hits left after hit cleaning, a multi-string
event contains at least six modules with recorded charges: three on
one string and three on another string. In these events, we reconstruct both the
zenith and the azimuth angles of the direction of a track. The second category is \emph{single-string} events that contain only one string in which modules have observed charges. Since all modules on a single string share approximately the same $x$ and $y$ coordinates, the azimuth angle of a track cannot be reconstructed. Example events for single-string and multi-string event fits are shown in Figs.~\ref{fig:santa-single-string-example} and \ref{fig:santa-multi-string-example}, respectively.

\subsection{Geometry of Tracks in IceCube}

\label{subsec:geometric-time-derivation}

To perform the $\chi^{2}$-fit on the observed hit times for a track hypothesis, we first need to derive the expected photon arrival time for an optical module at position $\vec{r}=(x,y,z)$ given the parameters of the hypothesis.

\begin{figure}[h]
\begin{centering}
\begin{tikzpicture}[scale=1,>=stealth]
	\path[name path=track] (0,3) -- (9,3);
	\node[shape=star,
	      star point height=1cm,
	      star point ratio=0.5,
	      draw, fill=black,
	      label=below:$\vec{p}(t_{\mathrm{em}})$] (emission) at (1,3) {};
	\draw[->, decorate,
	decoration={snake,amplitude=.4mm,segment length=2mm,post length=1mm}]
		(emission.center)
		-- node[sloped, above] {$d_{\gamma}$} +(40:4)
		node[label=above:DOM at $\vec{r}$] (dompos) {};
	\path[name path=cone] (dompos.center) -- +(-50:4);
	\draw[name intersections={of=track and cone, by=tip}]
		(dompos.center) -- node[sloped, above] {Cherenkov light cone} (tip)
		node[label=below:$\vec{p}(t_{\mathrm{geom}})$] (muonpos) {};
	\draw[fill=black, opacity=0.5] (dompos.center) circle (5pt);
	\draw[color=black, ->, style=very thick] (0,3) node[anchor=north]{muon} -- (muonpos.center);
	\draw (emission.center) +(1,0) node[anchor=south east]{$\theta_c$}  arc (0:40:1);
	\path (emission.center)
		-- node[shape=circle,
			fill=black,
			label=below:$\vec{q}$] (vertex) {}
		(tip);
	\draw[->] (vertex.center) -- node[sloped, below] {$\vec{r}-\vec{q}$} (dompos.center);
	\draw[->] (vertex.center) ++(0.2, 0.2) -- node[above] {$\vec{u}$} +(1,0);

\end{tikzpicture}\par
\end{centering}
\caption{\label{fig:Detailed-track-geometry}Detailed geometry of a light cone
created by a track. $\vec{q}$ is the position of the anchor point
and $\vec{r}$ is the position of the optical module. $\vec{p}(t_{\mathrm{em}})$
and $\vec{p}(t_{\mathrm{geom}})$ are the positions of the muon at
the time the photon is emitted and when it is geometrically expected
to arrive, respectively.}
\end{figure}
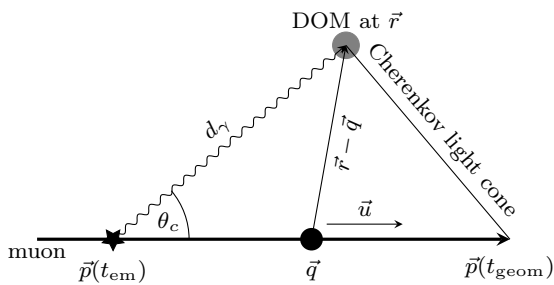

We characterize a track by a normalized direction vector $\vec{u}=(u_{x},u_{y},u_{z})$,
an anchor point $\vec{q}=(q_{x},q_{y},q_{z})$ and a time $t_{0}$
at which the particle passes through $\vec{q}$. In this simplified
hypothesis, tracks are modeled as being infinite in both directions;
there are no parameters to fix the start and end position and the
velocity is fixed to the vacuum speed of light, $c$. Since the reconstruction ignores DOMs that have not recorded any pulses, the fact that the true track length is finite only makes a negligible  difference.
Without scattering, all Cherenkov photons lie on a cone with an opening
angle $\theta_{c}$ (see Fig.~\ref{fig:Detailed-track-geometry})
whose tip is at the position of the particle at the time $\vec{p}(t)$. The opening angle satisfies $\cos(\theta_c)=1/n_{\mathrm{ph}}$, where $n_{\mathrm{ph}}$ is the phase index of refraction of the ice.

We solve the geometric equations analogously to~Ref.~\cite{Garza2014Measurement} assuming that a photon emitted by the moving particle travels in a straight line at a velocity of $c$ divided by the group index of refraction $n_{\mathrm{gr}}$, which gives the \emph{geometric time}, $t_{\mathrm{geom}}$, as a function of the track parameters

\begin{equation}
t_{\mathrm{geom}}=t_{0}+\frac{1}{c}\left(\left(\vec{r}-\vec{q}\right)\cdot\vec{u}+\frac{d_{\gamma}}{n_{\mathrm{ph}}}\left(n_{\mathrm{ph}}n_{\mathrm{gr}}-1\right)\right)\label{eq:t_geom-MS-track}
\end{equation}
where the distance traveled by the photon $d_\gamma$ is
\begin{equation}
d_{\gamma}=n_{\mathrm{ph}}\sqrt{\frac{1}{n_{\mathrm{ph}}^{2}-1}\left(\vec{u}\times\left(\vec{r}-\vec{q}\right)\right)^{2}}\,.\label{eq:photon-distance-3d}
\end{equation}

The group and phase indices of refraction depend on the wavelength, but for this reconstruction we use the values for the wavelength $\lambda=400\;\mathrm{nm}$\footnote{$400\;\mathrm{nm}$ is the wavelength of the highest acceptance of the optical modules.}, where $n_{\mathrm{gr}}=1.356$ and $n_{\mathrm{ph}}=1.319$~from Ref.~\cite{PRICE200197}.

\subsection{Fitting Procedure}
\label{sec:santa-loss}

\subsubsection{Loss function}

For a given set of parameters $\vec{\theta}=(\vec{u},\vec{q},t_0)$, we minimize a modified chi-square loss function given by
\begin{equation}
L(\vec{\theta})=\sum_{i=1}^{N}r^2_i
+
\frac{1}{\bar{q}}\sum_{i=1}^{N}\tilde{q}_i \frac{d_{\gamma,i}}{d_0}\,.\label{eq:chi-square-mod-loss}
\end{equation}
where $r^2_i$ is the squared time residual for each observed hit, $i$, between the observed time, $t_{\mathrm{obs}, i}$ and the geometric arrival time, $t_{\mathrm{geom},i}(\vec{\theta})$,

\begin{equation}
r_{i}^{2}=\left(\frac{t_{\mathrm{geom},i}(\vec{\theta})-t_{\mathrm{obs},i}}{\sigma_{t}}\right)^{2}\,.
\end{equation}

The uncertainty on the pulse time measurement is approximately $\sigma_{t}=3\,\mathrm{ns}$, corresponding to the readout rate of the modules~\cite{Abbasi:2008aa}.

The second term in eq.~\ref{eq:chi-square-mod-loss} is a regularization term that multiplies the distance traveled by a photon to the optical module that recorded it, $d_{\gamma,i}$, by the measured charge, $\tilde{q}_i$, to penalize solutions where a large charge is observed far away from the hypothesized track position. Because the modules are most sensitive on the side facing towards the bedrock, we correct the observed total charge in each DOM, $q_i$, for the sensitivity with
\begin{equation}
\tilde{q}_i=q_i\frac{2}{1+\cos(\vartheta_i)}\,,
\end{equation}
where $\vartheta_i$ is the angle between the direction of the photon
and a vector pointing up to the surface of the ice.
The parameter $d_{0}$ determines the relative contribution of the regularization term and is fixed to $7\,\mathrm{m}$. 

\subsubsection{Robust residual functions }
\label{sec:robust-losses}
After the hit selection described in Sec.~\ref{subsec:hit-selection}, a small number of hits from photons that have undergone significant amounts of scattering will remain that could strongly bias the fit result. We improve the robustness of the regression against such outliers by wrapping the squared residuals for each pulse in eq. \ref{eq:chi-square-mod-loss}, $r^{2}_i$, 
with the Cauchy robust loss function
\begin{equation}
r_i^2 \rightarrow \phi(r_{i}^{2})=\log\left(1+r_{i}^{2}\right)\,.
\label{eq:cauchy-loss}
\end{equation}
It reproduces the original $r^{2}$ residual
for small values of $r$, but grows more slowly than $r^2$ for large values of $r$, so that outliers are effectively given less weight.

Additionally, we choose the point of a “soft cut-off”,
denoted herein as $C$, at which the residual diverges from the regular $r_{i}^{2}$
in units of standard deviations by setting 
\begin{equation}
\phi(r_{i}^{2})\rightarrow\phi\left(\nicefrac{r_{i}^{2}}{C^{2}}\right)C^{2}\,.
\end{equation}

\begin{figure}[h]
\centering
\includegraphics[width=8cm]{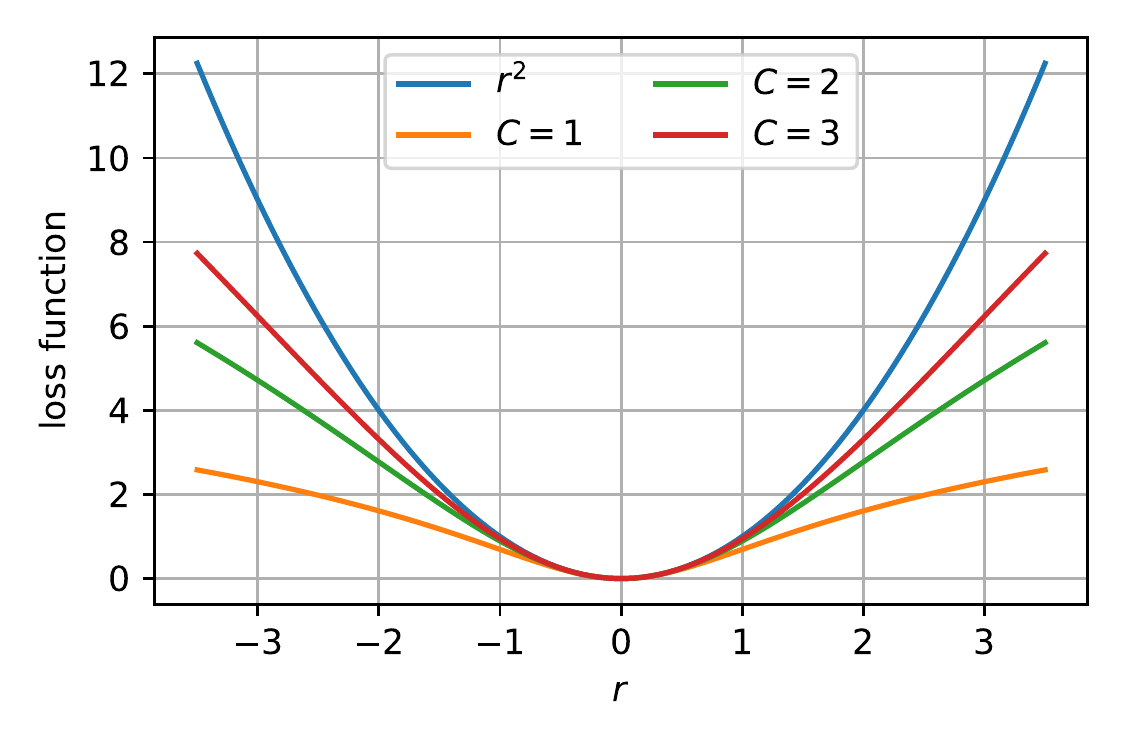}
\caption{\label{fig:robust-losses-example}The Cauchy robust loss function for different values of the scaling parameter, $C$. In a $\chi^{2}$
fit, the residual is the difference between the model and the observation
in units of standard deviations, $r=\frac{x-\mu}{\sigma}$ .}
\end{figure}

Figure~\ref{fig:robust-losses-example} shows the Cauchy loss function
for different values of $C$.  The choice of the value of $C$ is a trade-off. If it is too large, then the fit can be strongly influenced by single outliers. If it is too small, then the fit ignores too many hits and falls into degenerate solutions. We found the optimal value to be $C=3$. At this setting, the fit can effectively ignore hits that are far away from the Cherenkov cone.

As a further constraint on the regression, we apply the robust residual function from eq.~\ref{eq:cauchy-loss} only to pulses where the observed time is later than the expected photon time, since we expect that scattering would only cause photons to arrive too late, never too early. 

\medskip

The performance of the \textsc{santa} algorithm will be presented, together with the next algorithm discussed, in Sec.~\ref{sec:performance}.

\section{\textsc{retro} --- Embracing Scattered Light}
\label{sec:retro}

Light from neutrino interactions has a high chance of undergoing (multiple) scattering before being detected in IceCube.
The effective scattering coefficient for 400~nm light (peak acceptance of optical modules) in the region of DeepCore is about $b_e  \approx 0.03$-$0.04\,m^{-1}$ \cite{Ackermann:2006pva}, meaning an effective scattering length of 25-33\,m. This length is smaller than the inter-string spacing in IceCube of $\sim125$\,m, and DeepCore of $\sim45$\,m.
Consequently the geometrically propagated distance is enlarged, and the time of arrival of photons is delayed with respect to the geometrical propagation time (Eq.~\ref{eq:t_geom-MS-track}), and the photon direction will be diffused.
Thus, we developed the likelihood-based reconstruction method \textsc{retro}, which can handle scattered light and hence applies to all events.

\subsection{Likelihood}

In order to correctly model scattered photons in the event reconstruction, we derive the distribution $P(t|\vec{\theta})$, that describes the number of photons expected in a DOM as a function of arrival time $t$, given the neutrino event parameters $\vec{\theta}$. We will refer to $P(t|\vec{\theta})$ as the \textit{ unnormalized} probability density, in accordance with the language of Barlow~\cite{Barlow:1990vc}. These distributions encode the propagation of light through the inhomogeneous South Pole ice, and are not analytically available. Sections~\ref{sec:model} \& \ref{sec:tables} describe how these distributions can be approximated using \textsc{retro}---the \textit{reverse table reconstruction}.

Given the distributions, we can write down a likelihood featuring all registered photons at times $t_i$, based on an extended likelihood~\cite{Barlow:1990vc}:

\begin{equation}
    \log{\mathcal{L}(\vec{\theta})} = \sum_i \log{P(t_i|\vec{\theta})} - \Lambda(\vec{\theta}) - N
    \label{eq:llh}
\end{equation}
where $P(t_i|\vec{\theta}) = \lambda(t_i|\vec{\theta}) + n$ is the time dependent, unnormalized probability of registering photons. The term $\lambda(t_i|\vec{\theta})$ is the contribution from the actual neutrino event and will be introduced later, while the term $n$ represents a noise rate (see Ref.~\cite{Aartsen:2016nxy}), $\Lambda(\vec{\theta}) = \int_{TW}{\lambda(t|\vec{\theta})dt}$ is the total number of expected photons from the neutrino event over the trigger window (TW) and $N$ is the time-integrated noise rate. We can omit the last term $N$, as it is constant under variations of $\vec{\theta}$.
Since our reconstruction is based on a collection of pulses ${t_i}$ with charges ${q_i}$, as described in Sec.~\ref{sec:pulses}, we adapt the above expression to a charge weighted one: $\log{\mathcal{L}} = \sum_i q_i \cdot \log{P(t_i)} - \Lambda$. This log-likelihood is computed for all of the 5160 DOMs in IceCube and summed up.

\begin{figure}[h]
    \centering
    \includegraphics[width=\linewidth]{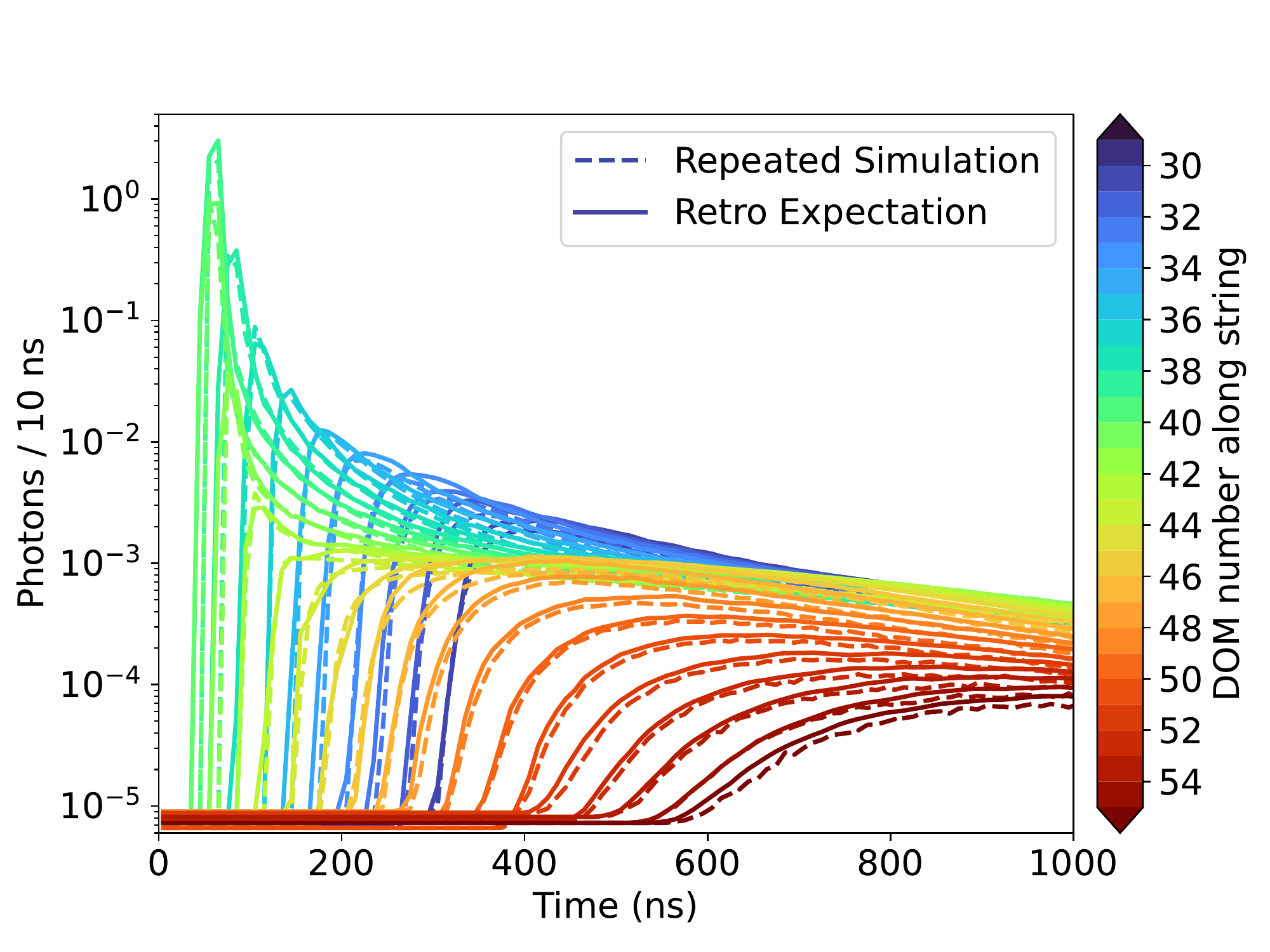}
    \caption{Expected photon time distribution in several DOMs along string 86, for a straight up-going muon with an energy 20 GeV, starting at $(0, 0, -350)$\,m in IceCube coordinates. The dashed lines show the average from $10^7$ repeated simulations, while the solid lines show the expected distributions based on the tables and event model discussed here.}
    \label{fig:time_dists}
\end{figure}

In simulations for IceCube, events are generated via Monte Carlo (MC) methods and the resulting photons are propagated via ray tracing \cite{Chirkin:2013tma}. This means that a simulated event corresponds to one particular random variate of the underlying probability density. Through repeated simulation, however, one can approximate the time distributions $P$.
An example of such distributions is shown in Fig.~\ref{fig:time_dists} for one particular choice of event parameters $\vec{\theta}$.
Such repeated simulations \cite{Chirkin:2013avz} allow for the reconstruction of single events while including the full level of detail available in IceCube simulation.
The main challenges for this approach are the large number of simulations necessary for creating the expected distributions and the associated computational cost, and the large number of events that need to be reconstructed. To address those, we will follow an alternative path to arrive at approximate distributions by splitting up the process into a parameterized event model (Sec.~\ref{sec:model}) and a tabulated detector response (Sec.~\ref{sec:tables}) discussed next.

\subsubsection{Parameterized Event Model}
\label{sec:model}

\begin{figure}[h]
    \centering
    \includegraphics[width=\linewidth]{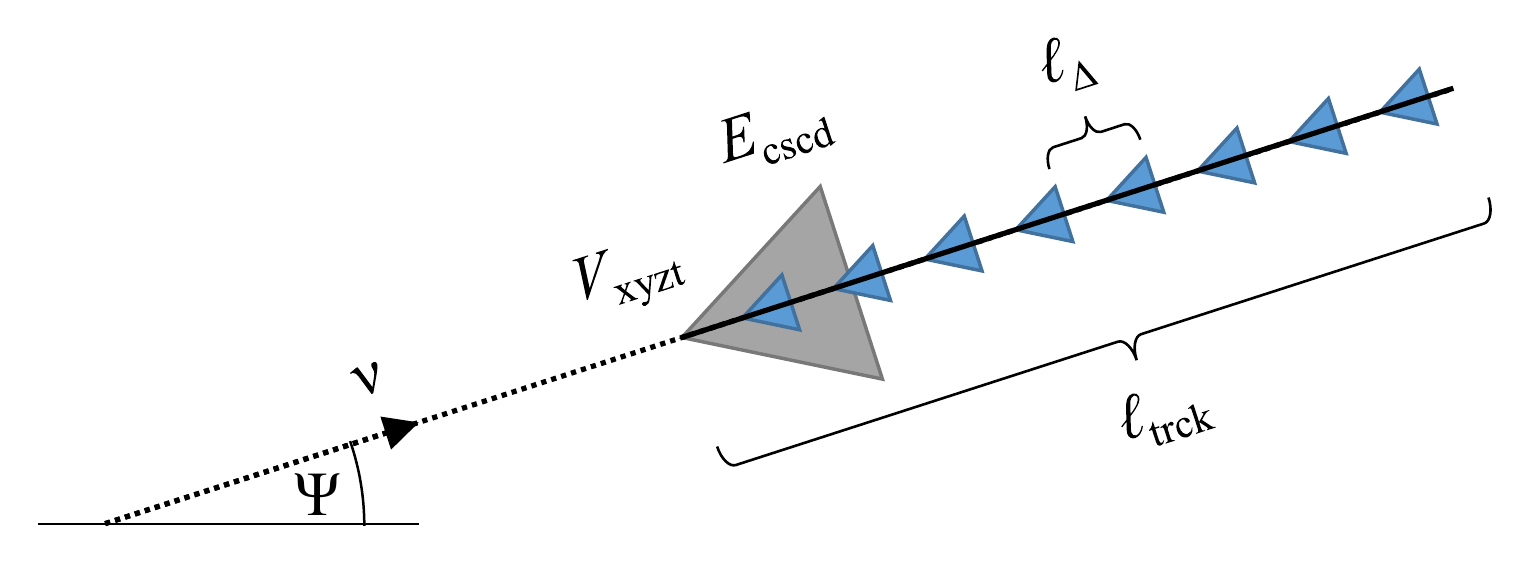}
    \caption{Sketch of the eight-dimensional event model for a neutrino $\nu$ incoming at angle $\Psi$ and interacting at vertex position and time $V_{xyzt}$. The cascade part is modeled as a single Cherenkov emitter (gray color) placed at the vertex in the same direction as the incident neutrino, and scaled by the expected number of photons per energy $\Ecscd$. The track part consists of $n$ (here $n=8$) emitters (blue color) placed at distances $\ell_\Delta$ along the track to reach the full length of \ltrck. }
    \label{fig:model}
\end{figure}

The parameterized event model describes how much Cherenkov light is emitted in the ice at different locations and times, given a set of event parameters: \begin{equation}
    \vec{\theta} = (\vertex, \Psi, \Ecscd, \Etrck) .
\end{equation}
This light emission is translated into a set of discrete Cherenkov-light emitters in the ice, which are then used together with the detector response to evaluate the likelihood function.
Our parameterization depends on the following eight parameters (see also Fig.~\ref{fig:model}):
\begin{itemize}
    \item \vertex: the neutrino's interaction point in ($x,y,z$) in the IceCube coordinate system and interaction time with respect to the trigger window
    \item $\Psi$: the zenith and azimuth angles $(\vartheta, \varphi)$ of the neutrino's direction of travel
    \item \Etrck: the energy carried by the outgoing muon in \numucc interactions, otherwise zero
    \item \Ecscd: all deposited energy in the interaction excluding \Etrck
\end{itemize}

The track part of the hypothesis is modeled as a series of colinear, constant luminosity Cherenkov emitters (2450 photons/m, \cite{Aartsen:2013rt}) placed every $\ell_\Delta = 3~\text{ns} \cdot c \approx 0.9$~m along the trajectory. Assuming constant energy loss, the track energy and length are related as $\Etrck / \textrm{GeV} \approx 11/50 \cdot \ltrck / \textrm{m} $ \cite{Aartsen:2013rt}.
The cascade part of the hypothesis is modeled as a single point emitting Cherenkov light source located at the vertex and co-directional with the muon track.\footnote{More detailed and realistic cascade light emission modeling have been studied, but did not significantly improve reconstruction performance.} Its light production is parameterized by 5.21\,m/GeV muon track equivalent ($\sim 12800$ photons per GeV) \cite{Aartsen:2013rt}.

These assumptions allow us to divide up the likelihood function into separate terms describing the cascade part with linear energy scaling, and the track part with a sum of discrete emitters every $\vec{\Delta}$ of length $\ell_\Delta$ along a line:

\begin{equation}
\begin{split}
    \lambda(t|\vec{\theta}) & = \lambda(t|\vertex, \Psi, \Ecscd, \Etrck)\\
    &=\lambda_{cscd}(t|\vertex, \Psi, \Ecscd)\\
    &\qquad+\lambda_{trck}(t|\vertex, \Psi, \Etrck)\\
    & \approx \Ecscd \cdot 12800 \cdot \lambda_{\gamma}(t|\vertex, \Psi)\\
    &\qquad+\sum_{k=0}^{\lfloor\ltrck/\ell_\Delta\rfloor} 2450 \cdot \ell_\Delta \cdot \lambda_\gamma(t|\vertex + k\cdot\vec{\Delta}, \Psi)
\end{split}
\label{eq:retro_llh}
\end{equation}
where the terms $\lambda_\gamma(t)$ describe the charge vs. time distributions for individual Cherenkov photons, and will be discussed in the next section.

This event model is chosen to obtain sufficient reconstruction accuracy while being fast enough to evaluate to make the analysis computationally feasible. Currently, the model does not account for subdominant stochastic losses along the muon track (which are generally small for our energies of interest \cite{Chirkin:2004hz}), the possibility that the cascade’s axis and track are not collinear, and longitudinal and off-axis shower development.
To partially correct for these shortcomings, the reconstructed energies undergo a post-reconstruction correction, described in Sec.~\ref{sec:reco_energy_proxy}.
More complicated and differently parameterized event models have been studied, but for the events of interest, no significant gain in reconstruction accuracy was found, while the computational load increased. These assumptions only hold for the low energy events targeted here, with energies in the range of around 10 -- 100 GeV.

\subsubsection{Detector Response}
\label{sec:tables}

The detector response characterizes the term \linebreak $\lambda_\gamma( t|(x,y,z,t)_\gamma,\Psi_\gamma)$, that expresses the probability of registering a photon emitted from a Cherenkov emitter at $(x,y,z,t)_\gamma$ in the direction $\Psi_\gamma$. These $\lambda$ terms encode the scattering and absorption in the ice, and are therefore location and orientation dependent. We are approximating the term using high statistics MC simulation to generate large lookup tables.
The tables are generated using the \textsc{CLSim} software\footnote{\url{http://github.com/claudiok/clsim}} provided with the Spice Lea ice model \cite{Chirkin:2013lpu} to simulate the propagation of photons in the South Pole ice.

\begin{figure}[ht]
    \centering
    \includegraphics[width=\linewidth]{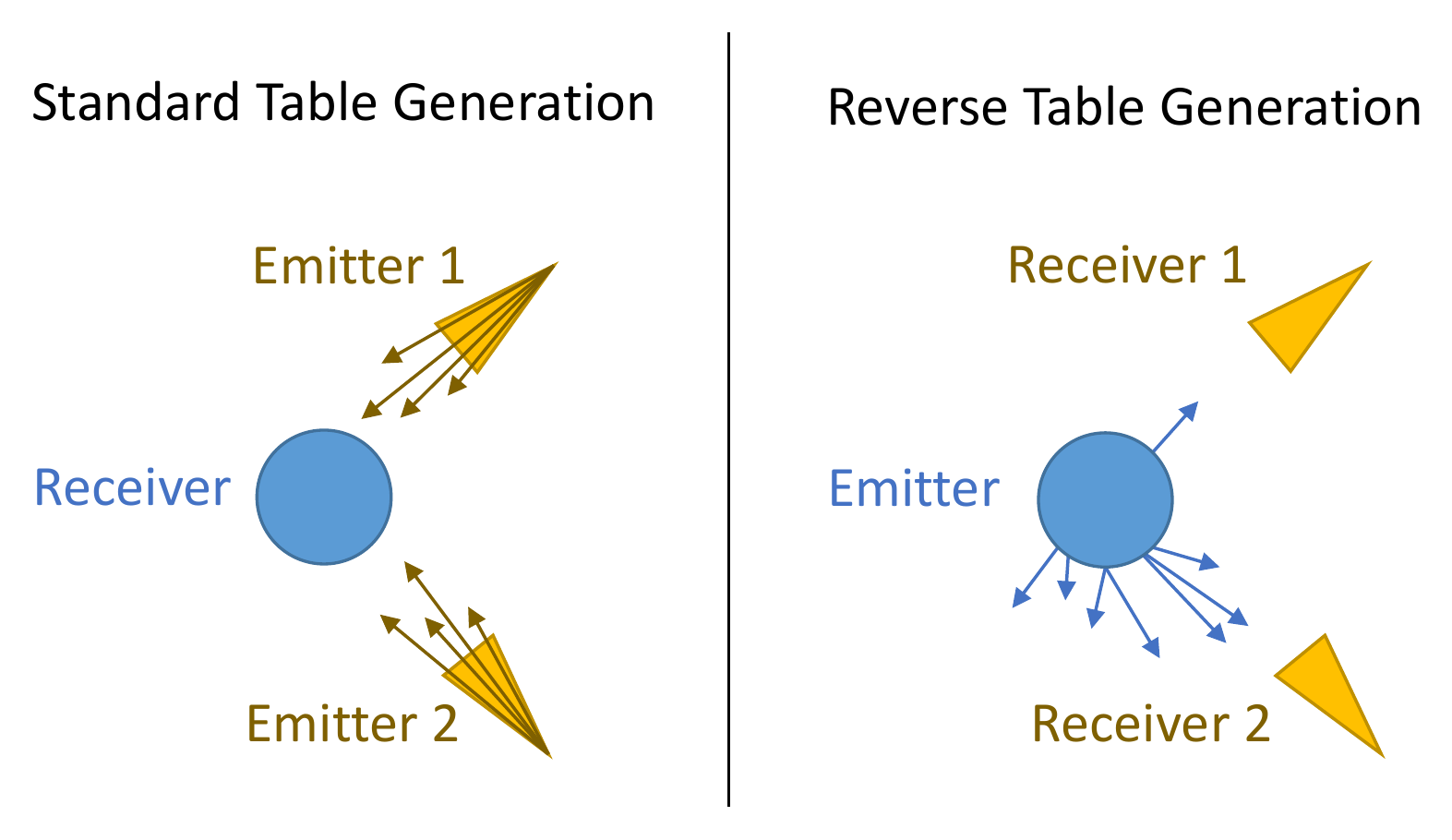}
    \caption{Illustration of photon tables generated in the standard approach (left) and the reverse \textsc{retro} approach (right). The circle in blue color denotes an IceCube DOM, and the triangles in yellow Cherenkov light sources.}
    \label{fig:retro}
\end{figure}

Other approaches based on tables, e.g. Refs.~\cite{Lundberg:2007mf} and \cite{Leuermann}, (referred to as ``standard'' approach in this paragraph) use Cherenkov-spectrum light sources placed at random points throughout the ice as if these are due to actual physical particles. Our tables, in contrast, treat PMTs as if they emit light. Assuming that absorption and scattering are time-symmetric processes, we reverse the role of the emitter and the receiver here (hence the name ``\textsc{retro}''). 
This difference is also illustrated in Fig.~\ref{fig:retro}, showing an example for the calculation of the photons expected in a DOM for two Cherenkov light sources at different positions and orientations. In the standard approach, two separate emitters would be simulated to represent both source hypotheses. In the \textsc{retro} reverse procedure, a single table with photons emitted from the DOM is needed to account for the different hypotheses.
Producing tables in each manner captures different symmetries and samples the space differently. In \textsc{retro}, for instance, the origin of the table is fixed at the DOM position---a position that is also fixed in the actual detector. In standard table, however, the origin is fixed at the source position, which is a variable of the reconstruction.
Another difference is due to the IceCube DOMs not having isotropic photon detection efficiency. They are mostly sensitive to photons coming in from below, i.e. towards the face of the PMT, and have almost no sensitivity to photons from the opposite side, i.e. the base of the PMT. In \textsc{retro} tables, photons are emitted according to this detection efficiency, meaning that only photons that can be observed are simulated. In contrast, standard tables generate typically similar amounts of photons for any emitter orientation, but emitters that point towards the PMT face will generate more PMT hits, while for emitters facing the other way many photons will never enter the detector acceptance (see Fig.~\ref{fig:retro}).
\textsc{retro} tables assume rotational symmetry about the DOM’s axis of symmetry, i.e. the ice is assumed to only change properties as a function of depth---an assumption that is only approximately true as the South Pole ice features an optical anisotropy~\cite{Chirkin:2013lpu}, and the ice layers exhibit a slight tilt~\cite{Chirkin:2013lpu}.

\begin{figure}[ht]
    \centering
    \includegraphics[width=\linewidth]{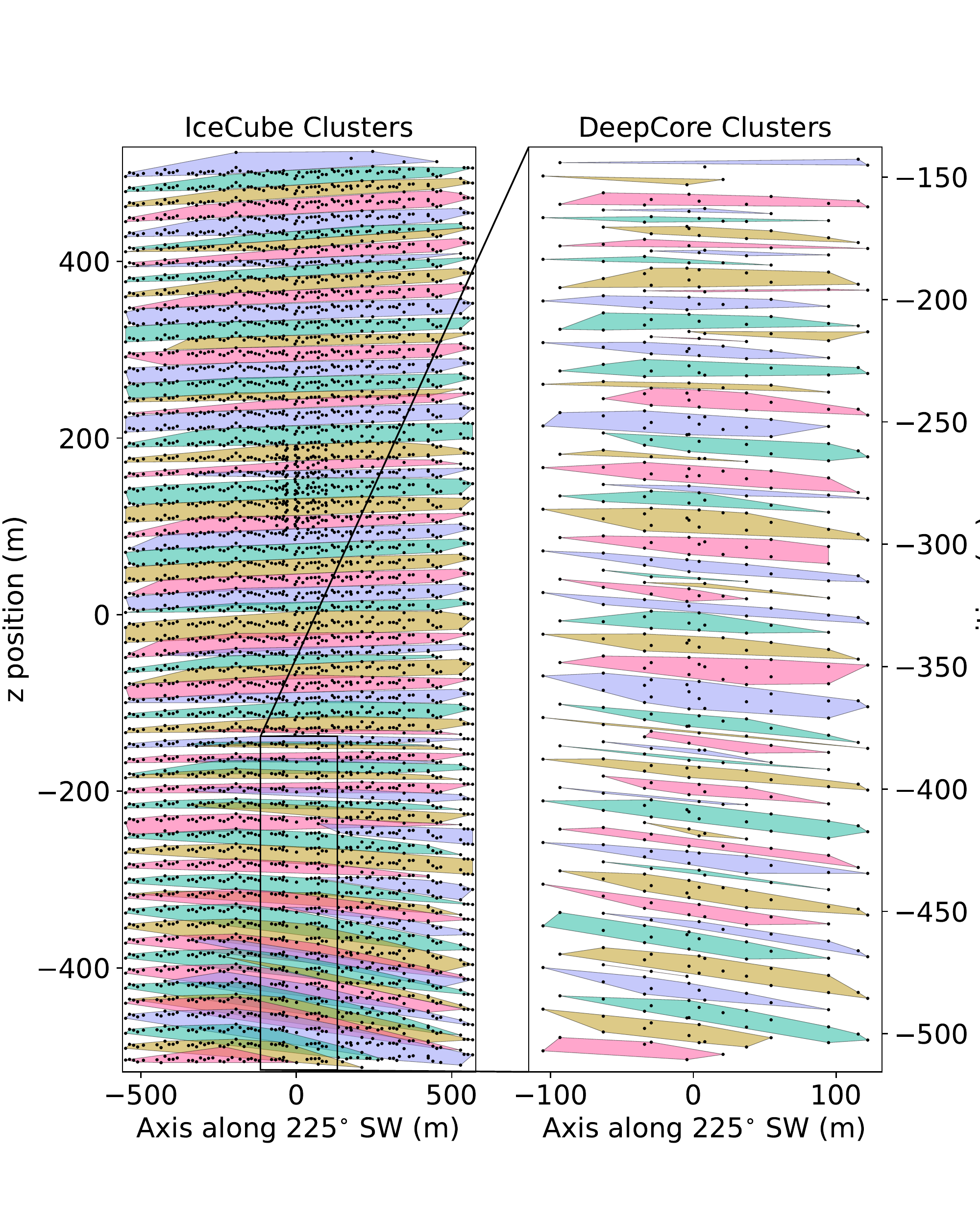}
    \caption{The grouping of DOMs into 140 separate ice property clusters via k-means, 80 for the larger IceCube volume (left), and 60 for the DeepCore volume (right). The plot shows the convex hull from DOM positions (black dots) in common clusters, and the varying colors are visually separating the clusters. At first order grouping happens in $z$-layers, while tilt behavior can be also be seen along the direction of the ice gradient of 225$^{\circ}$ SW.}
    \label{fig:clusters}
\end{figure}

We do not keep tables for each individual DOM, but merge similar ones (i.e. DOMs in regions of similar ice properties) together resulting in a total of 140 separate tables to represent the 5160 DOMs. This grouping is achieved via a k-means clustering algorithm \cite{macqueen1967some,lloyd1982least} and the resulting clusters are shown in Fig.~\ref{fig:clusters} for the IceCube and DeepCore volumes.
The resulting clusters reflect the structure of the South Pole ice properties, grouping DOMs in the depth-dependent and tilted ice layers together.

\begin{table}[h]
\centering
\begin{tabular}{|l|l|l|l|}
\hline
Dimension                & \# bins & Range               & Spacing    \\ \hline
$r$                      & 80      & $[0, 200]$~m   & $\sqrt{r}$ \\ \hline
$\Delta t$                & 100     & $[0, 4000]$~ns & $\sqrt{t}$ \\ \hline
$\cos{\vartheta}$        & 40      & $[-1, 1]$           & linear     \\ \hline
$\cos{\vartheta_\gamma}$ & 40      & $[-1, 1]$           & linear     \\ \hline
$|\Delta\varphi_\gamma|$ & 40      & $[0, \pi]$~rad & linear     \\ \hline
\end{tabular}
\caption{Table binning specifications for the five dimensions relative to the DOM position photons are emitted from. Spacing in $\sqrt{x}$ means quadratically increasing bin sizes in $x$.}
\label{tab:binning}
\end{table}

The photon tables are generated according to the binning specified in Table~\ref{tab:binning}, where $r$ is the distance from the center of the DOM to the photon’s location,
$\Delta t$ is the time of propagation in the ice minus $r/c'$, i.e. the time difference of the actual photon path minus the direct line of sight without scattering.
For the vector connecting the position of a photon back to the DOM it was emitted from, the quantity $\cos{\vartheta}$ is the cosine of the zenith angle of that vector.
The quantity $\cos{\vartheta_\gamma}$ is the cosine of the photon orientation zenith angle, and $\Delta\varphi_\gamma$ is the difference
between the photon orientation azimuth angle and the azimuth angle of the connecting vector mentioned just above. Because we assume azimuthal symmetry for the propagation of light, we can take the absolute
value $|\Delta\varphi_\gamma|$.
Photons are emitted from each DOM according to the average DOM zenith acceptance distribution \cite{Aartsen:2019tjl} and wavelength sensitivity distribution, where the latter
is weighted by the Cherenkov spectrum. Simulated photons are traced through the ice while being tabulated, i.e. accounting for which table bin volume ($V_{bin}$) they traversed ($N_{tab}$), normalized by the reference volume $V_{ref}$ that is the DOM surface area multiplied by the simulation step size.
Each DOM table is produced by simulating $N_{gen}=10^{10}$ photons.
In order to relate a number of emitted source photons ($N_{src}$) anywhere within the ice to the expected number in a DOM ($N_{exp}$), we use this table by applying correct normalization and looking up the number of tabulated photons ($N_{tab}$) in the bin where the source lies with respect to the DOM:

\begin{equation}
    N_{exp} = N_{src} \cdot \frac{N_{tab}}{N_{gen}} \cdot \frac{V_{ref}}{V_{bin}}
\end{equation}

Since the photons of interest (originating from neutrino events) are emitted from Cherenkov radiation, the generated tables are convolved with the angular emission profile (see Fig.~\ref{fig:Detailed-track-geometry}), such that the new directional angles in the table correspond to that of the charge generating the Cherenkov light cone (see also Fig.~\ref{fig:table_compression}).

\subsubsection{Table Compression}

A raw table, based on the above specified binning in five dimensions, contains $5.12\cdot 10^8$ bins, and we generate 140 different tables. In single precision floating points, the required space amounts to about $0.29$\,TB of memory---a value exceeding the available RAM in typical compute nodes available to us.
For this reason, we compress the raw tables with a custom procedure that is described below down to about $315$\,MB, achieving a compression factor of three orders of magnitude.

Since the directional distributions of photons follow the scattering behaviour in the ice, many of these distributions look similar. The full set of 140 tables can be split up into $(140 \times 80 \times 100 \times 40)$ bins of magnitude, and corresponding directionality distributions of $(40 \times 40)$ bins. We now replace the total 44.8 million directionality distributions with a collection of 4000 templates that are representive of the original distributions. This means that instead of the full table, we only store the magnitude (= number of photons) and template index. 

To generate a representative template library, we first linearly transform the distributions (i.e. the bin values of the distributions represented as 160-dimensional vectors) using a principal component analysis (PCA), and perform a k-means clustering based on the first 80 components to group similar distributions together, resulting in 4000 clusters. Then we average the original distributions of the members of each cluster into a template, resulting in 4000 template distributions. In the final step, the original tables are then replaced by finding the best matching template for each bin, i.e. the statistically most compatible template determined via the Pearson $\chi^2$-distance.

\begin{figure*}[ht]
    \centering
    \includegraphics[width=0.8\textwidth]{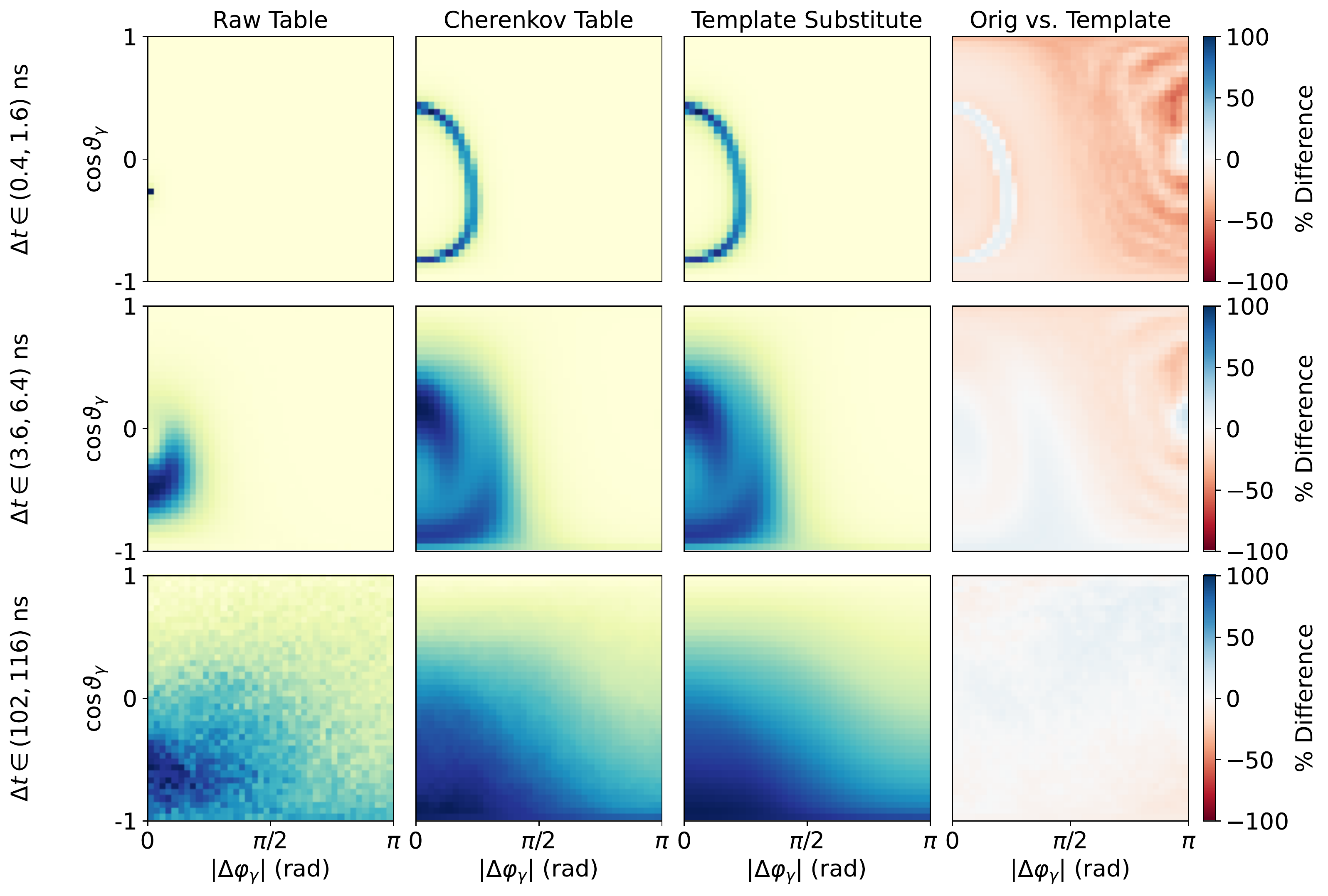}
    \caption{Example \textsc{retro} table slices showing the photon directions for $\Delta t$ bins (0.4,  1.6), (3.6,  6.4), and (102, 116)\,ns from top to bottom, at $r$-bin (7, 8)\,m and $\cos{\vartheta}$-bin (0.25, 0.3).
    The left column shows the photon direction distribution from emitted photons, the second column has the Cherenkov-cone folding applied, the third column shows the template substitute, and the last column shows the error introduced by the template compression. The distributions in the first three columns are normalized, and darker areas indicate higher photon density (linear scale).
    The first row contains almost exclusively direct (unscattered) photons, and hence the left most plot in that row has almost all photons in the bin of direct line of sight, and the Cherenkov table contains a sharp ring. The subsequent rows are at later times and contain more and more scattered light resulting in diffusion.
    The last column showing the percentage difference between compression and original exhibits the largest deviations in the right half, where light is back-scattered and statistics are very low.}
    \label{fig:table_compression}
\end{figure*}

Figure~\ref{fig:table_compression} shows a few example directionality distributions, their template substitutes, and the difference between the two. The fidelity of the compression is high and does not introduce unwanted artifacts. In the figure, larger deviations can be observed in the region $|\Delta\varphi| > \pi/2$, where photons are back scattered and hence very low photon statistics are present.  In this region, the templates offer smoothed-out distributions due to the averaging.

\subsubsection{Likelihood Landscape \& Comparison to Simulation}

Our parameterized event model and the compressed tables for the detector response---including all the approximations discussed---offer a sufficiently precise description of the reconstruction likelihood.
The expected number of photons vs. time, given a set of event parameters, is approximated well over several orders of magnitude.
This is illustrated in Fig.~\ref{fig:time_dists}, where we show the expectation from our model for the same event as the repeated simulations.  The expected photon counts span over five orders of magnitude and the time range displayed is one microsecond. An accurate description with deviations at the level of $\mathcal{O}(10\%)$ is achieved, and the overall shape of the curves with rising edge and diffusion tail are matched well.
The repeated simulations shown in Fig.~\ref{fig:time_dists} include the full complexity of event modeling available in IceCube simulation, and our simplified, parameterized model provides an overall reasonable approximation of these distributions.

\begin{figure*}[ht]
    \centering
    \includegraphics[width=0.75\textwidth]{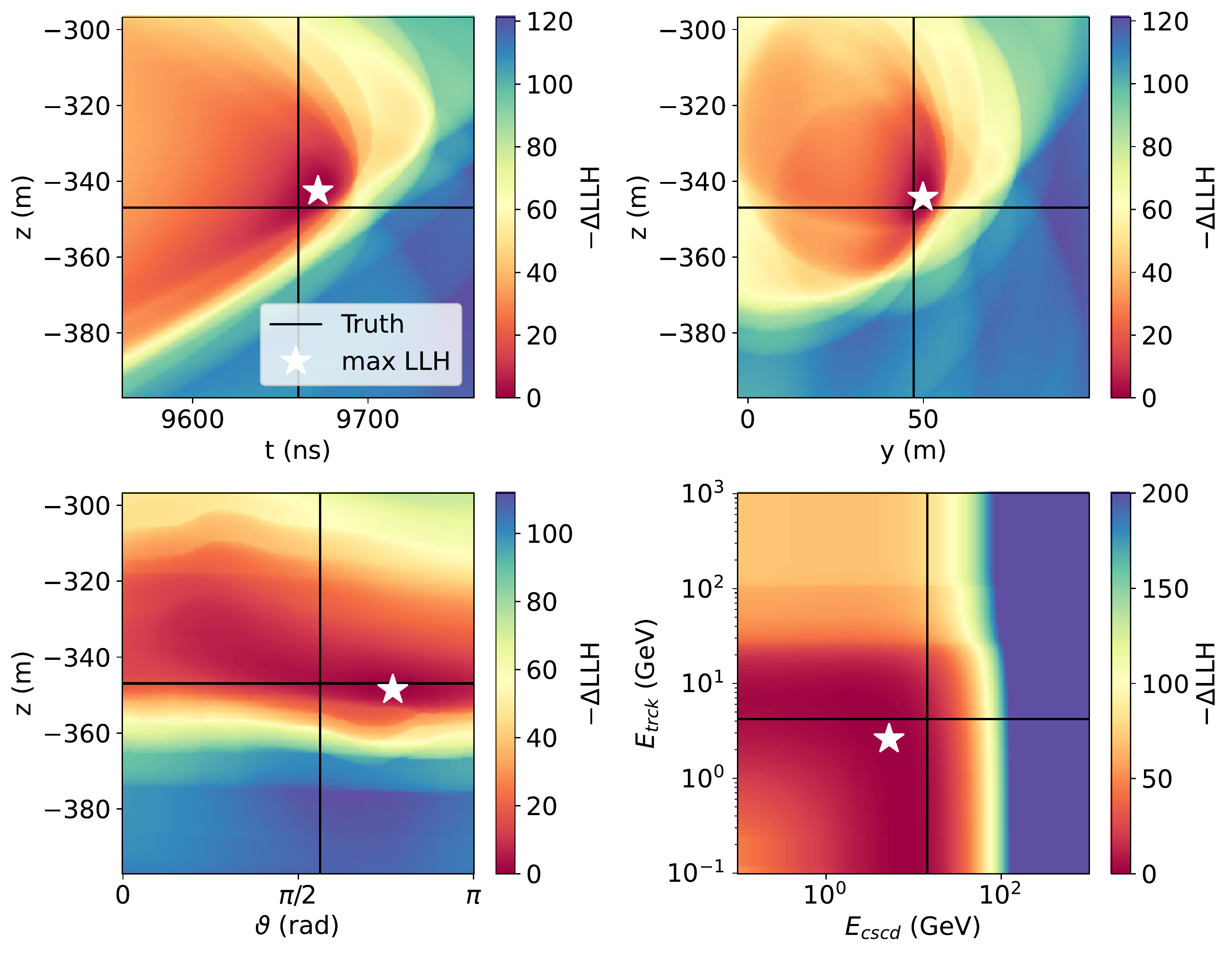}
    \caption{Slices through the likelihood landscape, with the remaining six event parameters set to truth in each panel. The white star indicates the maximum likelihood point within the slice. The true values (shown as black crosshairs) for the depicted \numucc example event are $V_{xyzt}$ = (8.0\,m, 47.2\,m, -346.9\,m, 9660\,ns), $(\varphi, \vartheta)$ = (1.65, 1.77)\,rad, and deposited energies (\Ecscd, \Etrck) = (14.3, 4.2) GeV.}
    \label{fig:llh}
\end{figure*}

Our reconstruction likelihood $\mathcal{L}(x|\vec{\theta})$ (Eq.~\ref{eq:retro_llh}) is, in general, not differentiable, as it depends on discrete table entries, and those are in addition susceptible to statistical fluctuations.
The likelihood exhibits a multimodal structure and is highly non-Gaussian. Figure~\ref{fig:llh} shows a few example slices through the likelihood of a typical event in two dimensions to illustrate its structure.

\subsection{Optimization}

Finding the maximum likelihood estimators $\hat{\vec{\theta}}$ for our reconstruction parameters requires a maximization of the likelihood function (Eq.~\ref{eq:llh}). We split the model parameter space up into the two energy terms (\Etrck\ and \Ecscd) that are maximized separately (see Sec.~\ref{sec:emin}) for any choice of the remaining six parameters that are maximized in the external loop described in Sec.~\ref{sec:dirmin}.

\subsubsection{Energy Optimization}
\label{sec:emin}

Since the cascade energy \Ecscd\ in our likelihood is realized via a simple scaling factor (see Eq.~\ref{eq:retro_llh}), the change and the derivative of the likelihood with respect to this parameter are known. To exploit this fact, we maximize the likelihood, given the remaining 7 parameters, via a binary search for the root of this derivative $\partial \mathcal{L} / \partial \Ecscd$. This provides the optimal cascade energy configuration as a function of the other 7 parameters.

For the track energy parameter \Etrck, the same does not apply, as the likelihood is a sum of independent track energy contributions summed along the extension of the length of the track (see also Eq.~\ref{eq:retro_llh}). For higher energies, more elements are added to extend the track, while the previous track segments remain unchanged. 
Based on this, we maximize the likelihood as a function of the track length, starting at \ltrck=0 (i.e. \Etrck=0), and adding increments, one at a time, to find the optimal solution. The cascade energy is re-optimized for every track configuration. This procedure is computationally advantageous, since the likelihood of all previous track segments that have already been computed can be reused and only the new increments need to be computed and added.

The likelihood difference between the \Etrck=0 and optimal \Etrck\ is also stored as a useful variable to distinguish track from cascade events (as used in Sec.~\ref{sec:pid}).

\subsubsection{Vertex \& Direction Optimization}
\label{sec:dirmin}

For the optimization of the remaining 6 parameters, we use a custom, derivative free, global optimization algorithm. 
It is based on the controlled random search with local mutation (\textsc{crs2}) algorithm described in \cite{Kaelo2006}, and extended to correctly treat angles (azimuth and zenith).
An open-source implementation is provided\footnote{\url{https://pypi.org/project/spherical-opt}} and more details can be found in \ref{sec:crs}. The customized treatment of spherical quantities (azimuth and zenith angles) is crucial for convergence and thus the performance of our reconstruction.

The two energy parameters are a nested optimization inside this outer 6-d optimization for vertex and direction, i.e. for any point in the 6-d vertex-direction space, the two energies are always optimized internally.

\paragraph{Seed Points}

The optimization is started by choosing 160 quasi-random seed points that are distributed according to a fast seed reconstruction ``\textsc{spefit}'' (see for example Ref.~\cite{Ahrens:2003fg} for details) with a spread of the points following the seed reconstruction’s resolution. The resolution of the \textsc{spefit} seed reconstruction is also shown in Sec.~\ref{sec:comparisons_angular_resolution}.

\paragraph{Convergence}

After each internal iteration of the algorithm, we evaluate four stopping criteria. If any of these criteria is met, the minimization is terminated. The following criteria are classified to have converged successfully:
\begin{enumerate}
   \item  Number of consecutive iterations without finding an improved point $\ge 1000$, or
   \item  Standard deviation of the log-likelihood values of all 160 current minimizer points $\le 0.5$, or
   \item  Standard deviation of the vertex position \vertex\ of all 160 current minimizer points $\le (x, y = 5\,\text{m}, z = 4\,\text{m}, t = 20\,\text{ns})$
\end{enumerate}
Termination by the last criterion indicates the minimization was unsuccessful:
\begin{enumerate}
   \setcounter{enumi}{3}
   \item  Number of iterations $\ge 10000$
\end{enumerate}

For a typical selection of IceCube DeepCore events for atmospheric neutrino oscillation studies (as used in Sec.~\ref{sec:performance}), we observe the optimization to terminate on average after $\langle n \rangle = 2229$ iterations of the optimization routine, while 99\% of events finish with less than 5305 iterations, and for all events the optimization converged successfully. With 11.3\% of cases stopped via condition 1, 39.1\% via condition 2 and 49.7\% of the cases terminating via condition 3.

\subsection{Energy Corrections}
\label{sec:reco_energy_proxy}
\subsubsection{Cascade Energy Conversion}

\begin{figure*}[t]
    \centering
    \includegraphics[width=0.95\textwidth]{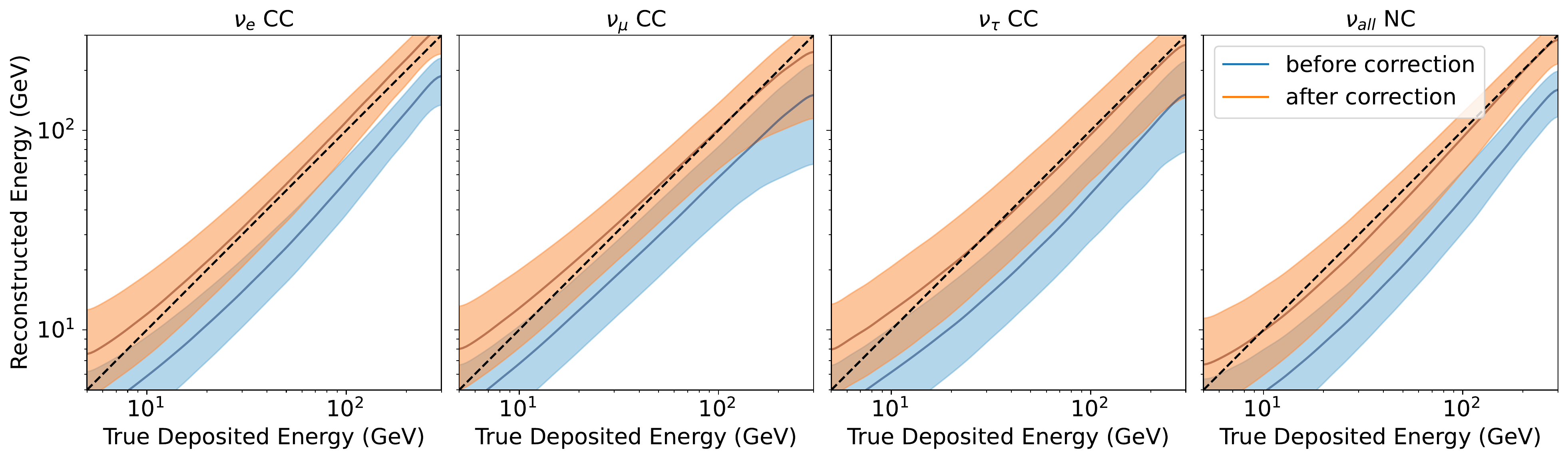}
    \caption{Reconstructed energy before and after the bias correction is applied. This bias correction is applied to make the reconstructed energy a better estimator of true deposited energy. The shaded region shows the 68\% containment regions, and the solid line shows the median. The dashed, black line is the 1:1 line where reconstructed energy exactly matches the true deposited energy. The correction pulls the median to more closely follow the 1:1 line.}
    \label{fig:energy_rescaling}
\end{figure*}

\textsc{retro} provides two energy estimates, \Ecscd\ and \Etrck, as described in Section \ref{sec:model}.
We reconstruct the energy as visible, electromagnetic-equivalent energy. While topological differences in the experimental signatures of hadronic and electromagnetic cascades are small, the true energy needs to be rescaled accounting for the proper hadronic photon yield.

To account for this, a correction factor $F$ is applied to convert the reconstructed EM cascade energy to a more accurate cascade energy estimate. As a proxy for light yield, the cumulative Cherenkov track length $T$ is used, which is found by summing all charged shower products with energies above the Cherenkov threshold. $F$ is defined to be the ratio of Cherenkov track lengths for hadronic and electromagnetic cascades for a given primary energy:

\begin{equation}
F(E)=\frac{T_{Hadr}(E)}{T_{EM}(E)}
\end{equation}

We use the functional form for $F$ described in \cite{Kowalsi_thesis} and \cite{Dunkman_thesis}:
\begin{equation}
F = F_{EM} + f_0(1 - F_{EM})
\end{equation} where $f_0$ is the relative Cherenkov activity of the pure hadronic portion of the cascade and $F_{EM}$ is the fraction of the total energy in the pure electromagnetic portion of the cascade. The following expression is used for $F_{EM}$ originating from \cite{hadronic_cascades}: 
\begin{equation}
F_{EM} = 1 - (E/E_0)^{-m}
\end{equation}
where $E_0$ and $m$ are model parameters which depend on the primary hadron and the detector material. The following parameter values are used: $E_0=0.188$ GeV, $m=0.163$, $f_0=0.310$ \cite{Dunkman_thesis}.

\subsubsection{Track Energy Conversion}

The second energy estimate from \textsc{retro}, \Etrck, is directly related to the optimal track length found during the reconstruction process. During the reconstruction, a constant energy loss of 0.22 GeV/m is assumed, as mentioned in Sec.~\ref{sec:model}. However, this is not a perfect assumption and we can get an improved estimate of the real track energy by using more realistic energy losses which are dependent on the primary muon energy. The energy of the track (i.e. the original muon) is recalculated by using an interpolation of the muon ranges and muon corresponding primary energies found in Table II-28 from \cite{GMS}.

\subsubsection{Bias Correction}
\label{sec:bias_correction}

With the above corrections applied, the resulting mean reconstructed cascade energies and track lengths still do not perfectly match their true parameter values. A linear scaling factor is applied to make the reconstructed parameters better estimators of the true deposited energies. A scaling factor of 1.7 for all cascade energies after being converted from EM to hadronic, and a scaling factor of 1.45 for all track lengths before being converted to track energy are used here. The total reconstructed energy is then the sum of these resulting two quantities. Figure \ref{fig:energy_rescaling} shows the total reconstructed energy versus the true deposited energy with and without these correction factors applied.

\subsection{Particle Identification (PID)}
\label{sec:pid}

To distinguish between track-like and cascade-like events in DeepCore, the likelihood ratio $\mathcal{L}(\hat{E}_\textrm{trck}) / \mathcal{L}(\Etrck=0)$ alone would make the most powerful classifier (``Neyman–Pearson'' lemma) given a perfect likelihood description. However, with the approximations in the likelihood discussed earlier, we find that a multivariate classifier with additional inputs to the likelihood ratio leads to an improved classification.
Here we use a boosted decision tree (BDT) based on the the \textsc{XGBoost} algorithm \cite{Chen:2016}.
The variables used as input features are the following reconstructed quantities determined by \textsc{retro}:
\begin{itemize}
    \item $\mathcal{L}(\hat{E}_\textrm{trck}) / \mathcal{L}(\Etrck=0)$: Likelihood ratio between the full reconstruction and a cascade-only hypothesis in which \Etrck=0, as mentioned in Section \ref{sec:emin}. 
    \item \ltrck: The reconstructed track length is a straightforward measure of how track-like an event is. A long reconstructed track indicates the presence of a muon track in the event.
    \item \Ecscd: \textsc{retro}'s estimate of how much light is in the cascade portion of the event (as opposed to the track component). 
    \item Zenith angle: The ability to identify tracks is zenith-dependent due to the geometry of the detector, because the spacing between adjacent DOMs in the vertical direction is smaller than in the $x-y$ plane. Therefore tracks that are nearly vertical are easier to identify than horizontal ones.
    \item Zenith Spread: This quantity is calculated from the distribution of points traversed by the minimizer before convergence. Tracks are typically associated with better pointing resolution due to their longer lever arm, and we expect events with a smaller zenith spread to be more track-like.
\end{itemize}

The classifier is trained on a selection of simulated \numucc and \nuecc with reconstructed energy between 5 and 500 GeV. For training, events are weighted with an unoscillated spectrum according to the HKKM flux \cite{Honda:2015}.
Sample balancing is performed for the relative contributions between classes, with an overall equal weight for true tracks and cascades during training. The sample is divided and 50\% of events are used for training, and 50\% for testing the model.

\begin{figure}[h]
    \centering
    \includegraphics[width = 0.9\linewidth]{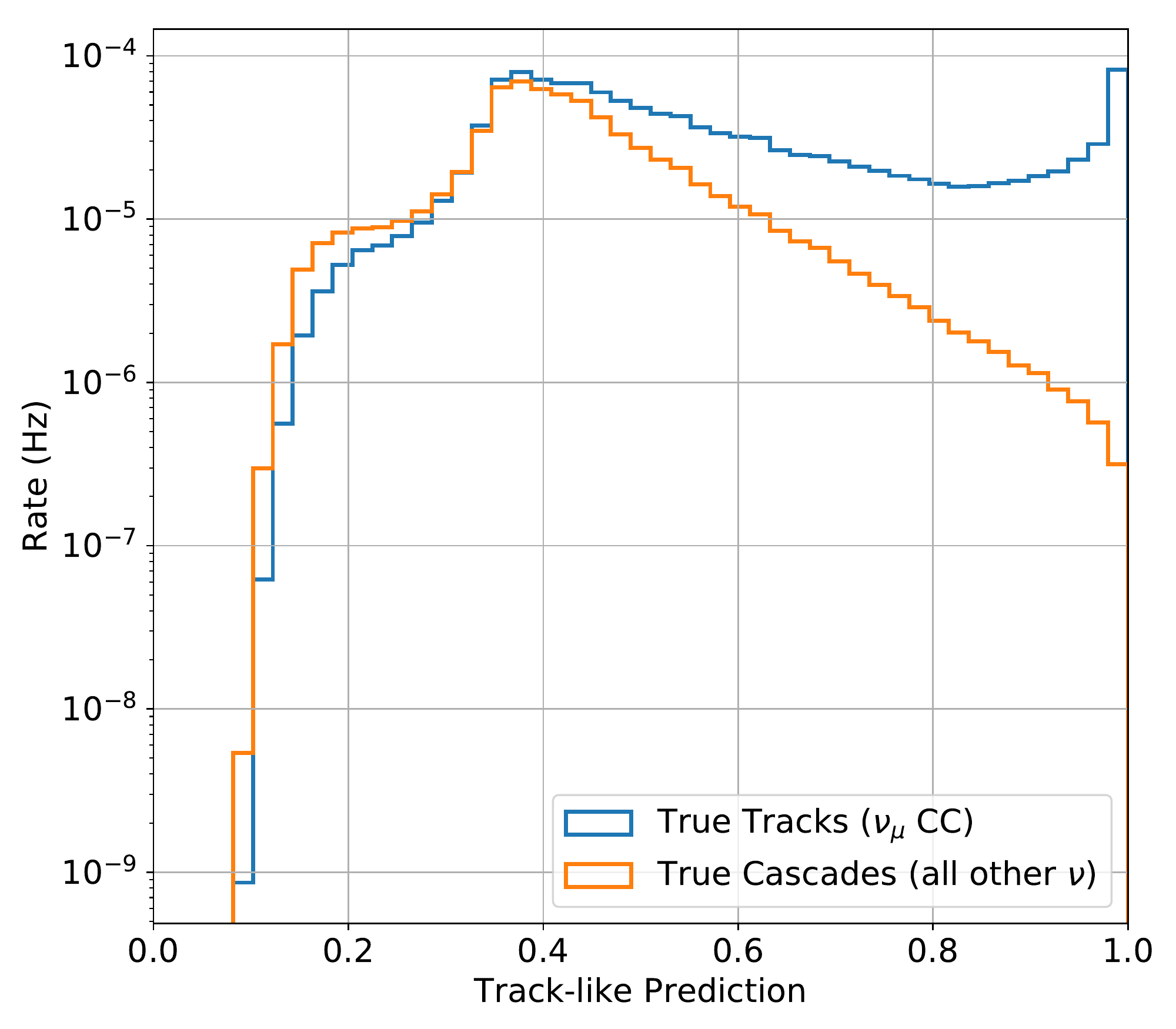}
    \caption{Distribution of PID scores for events that are tracks (blue) and cascades (orange).}
    \label{fig:pid_scores}
\end{figure}

\begin{figure}[h]
    \centering
	\includegraphics[width = 0.9\linewidth]{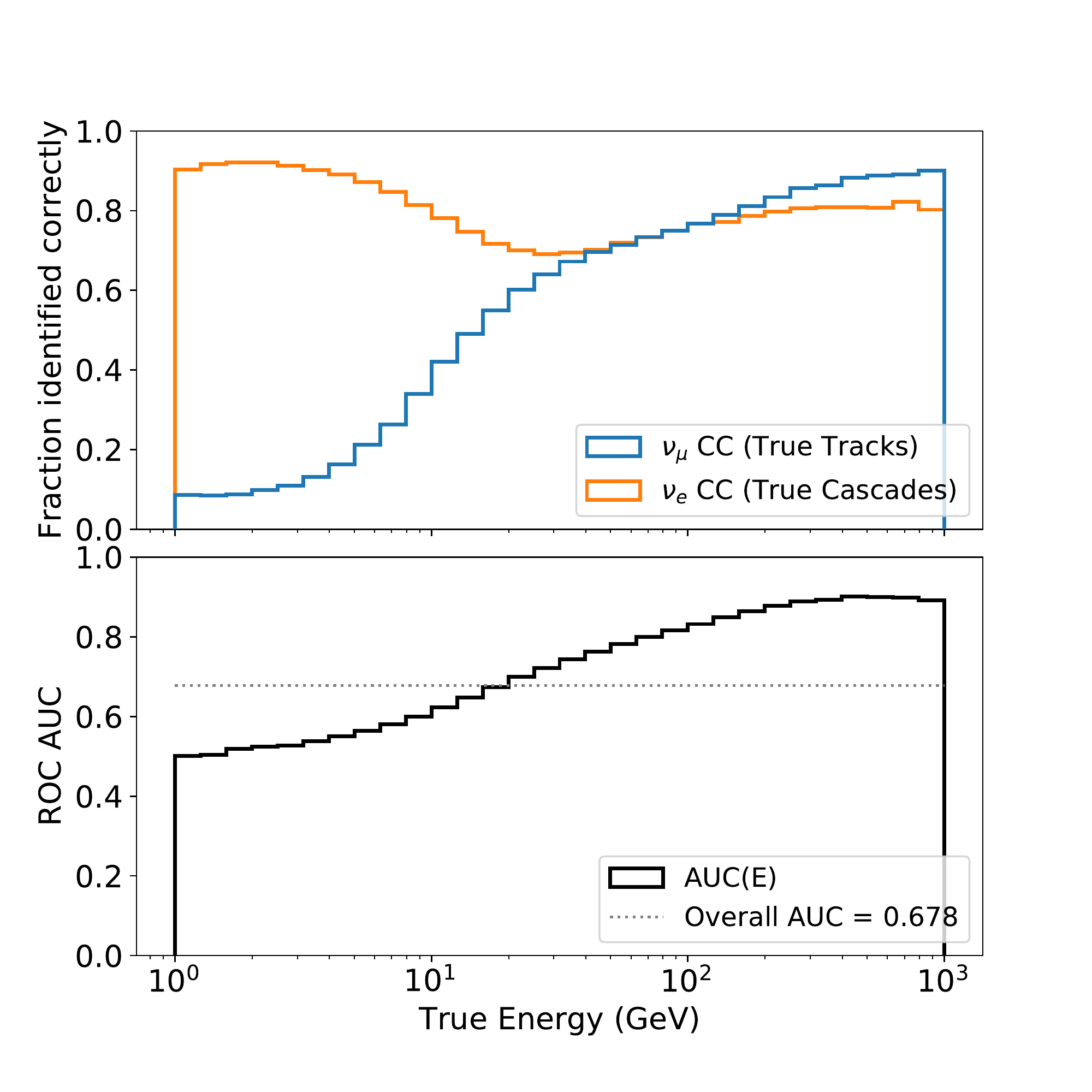}
	\caption{The top panel shows the fraction of events classified correctly as a function of energy. The bottom panel shows the area under the curve (AUC) for the receiver operating characteristics (ROC) for each energy band. The classifier performance is better at higher energies, while towards the lowest energies the AUC converges to 0.5, i.e. zero classification performance.}
	\label{fig:fractions_track_cascade}
\end{figure}

The output score of the BDT is a number between 0 and 1 indicating how track-like an event is, with 1 being the most track-like. Figure \ref{fig:pid_scores} shows the distribution of classifier scores for tracks and cascades for an example sample (see Fig.~\ref{fig:true_dists}). There is a peak at 1 in the track distribution as we expect, and cascades appear mostly at lower scores. Many events in the confusion region around the middle tend to be low energy events, and therefore have very few hits in the detector and consequently not enough information for the BDT to distinguish between event types.

Using an example cut of 0.5 to evaluate the performance of the classifier as a function of energy, we find that at the lowest energies almost everything is classified as a cascade and at the highest energies, the vast majority of events are classified correctly (see Fig.~\ref{fig:fractions_track_cascade}).

\section{Performance}
\label{sec:performance}

Finally, we demonstrate how the two reconstructions presented in this paper perform compared to each other for a common set of events. We also show the \textsc{retro} performance on a larger event sample, which includes events that do not pass the \textsc{santa} selection criteria.

To illustrate the algorithm's performance, we use a neutrino-only MC set simulated with primary neutrino energies between 1\;GeV and 10\;TeV, following an $E^{-2}$ spectrum, and containing about 8.1 million events passing a selection process similar to what is used in Ref.~\cite{IceCube:2019dyb}. In all following figures (except those showing reconstruction time) the events are weighted by the HKKM flux\cite{Honda:2015}, oscillation probability calculated with the \textit{NuFIT} 2.2\cite{Gonzalez-Garcia:2014bfa} best-fit values, and interaction and detection probabilities to arrive at a physical spectrum close to what is expected from real data. The true parameter distributions of this sample (with selection and weights applied) are shown in Fig.~\ref{fig:true_dists} in the appendix.\\
About 3.2 million events fulfill the criteria to be reconstructed with \textsc{santa}. In the following plots that compare \textsc{retro} and \textsc{santa} results (see Figs.~\ref{fig:angular_res}, \ref{fig:coszen_reco_true}, \& \ref{fig:santa_time}), only those events are used. The plots showing \textsc{retro} alone (see Figs.~\ref{fig:retro_time} \& \ref{fig:full_retro}) use the full event sample.

\subsection{Comparison between reconstructions}
Since \textsc{santa} only estimates the particle direction, we compare the azimuth and zenith resolutions of the two reconstructions. 
In addition, we also compare the time needed to perform a fit.

\subsubsection{Direction angles}
\label{sec:comparisons_angular_resolution}
\begin{figure}[ht]
\centering
\begin{subfigure}{.48\textwidth}
  \centering
  \includegraphics[width=\linewidth]{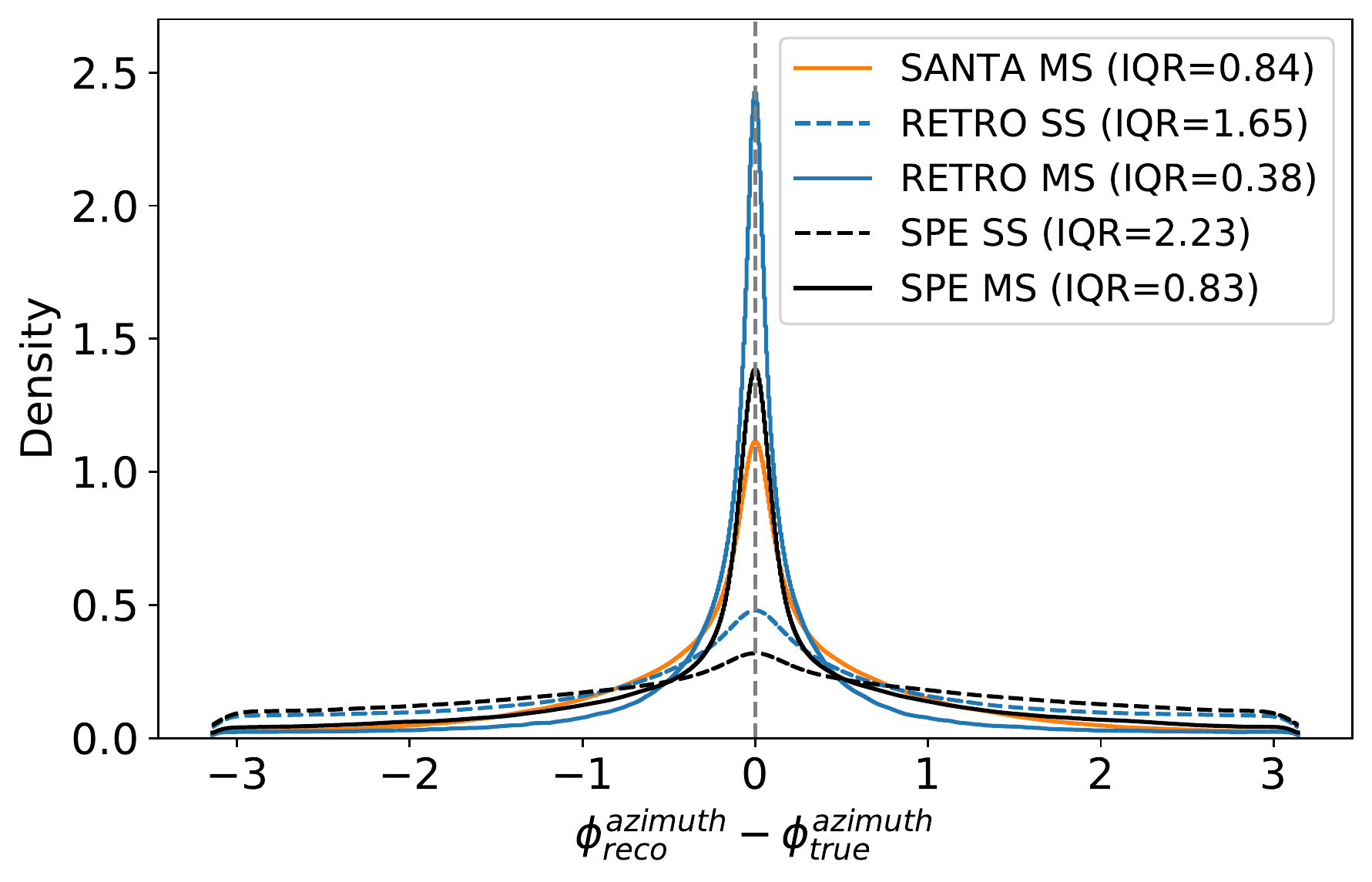}
\end{subfigure}
\hspace{0.01\textwidth}
\begin{subfigure}{.48\textwidth}
  \centering
  \includegraphics[width=\linewidth]{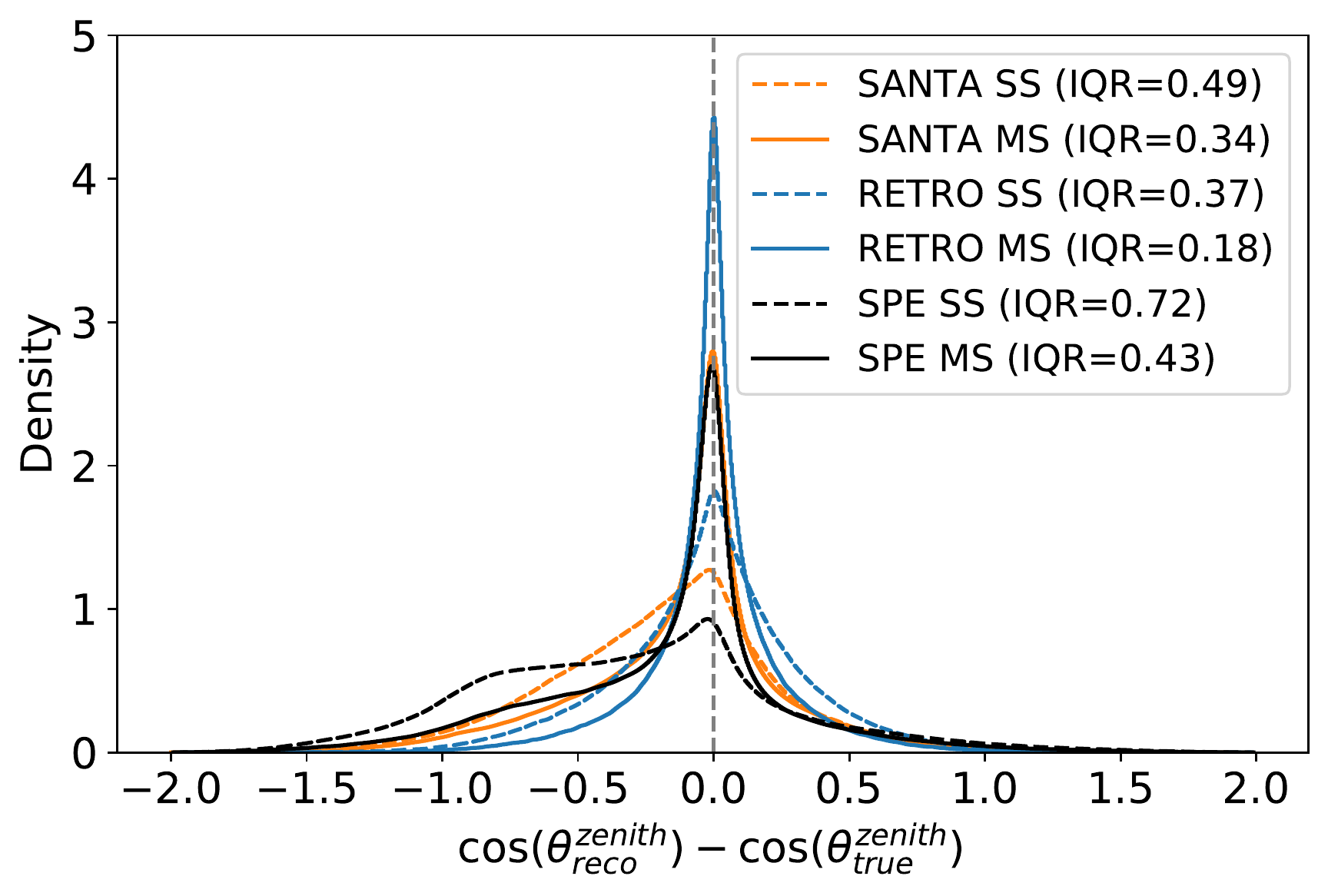}
\end{subfigure}
\hspace{0.01\textwidth}
\begin{subfigure}{.48\textwidth}
  \centering
  \includegraphics[width=\linewidth]{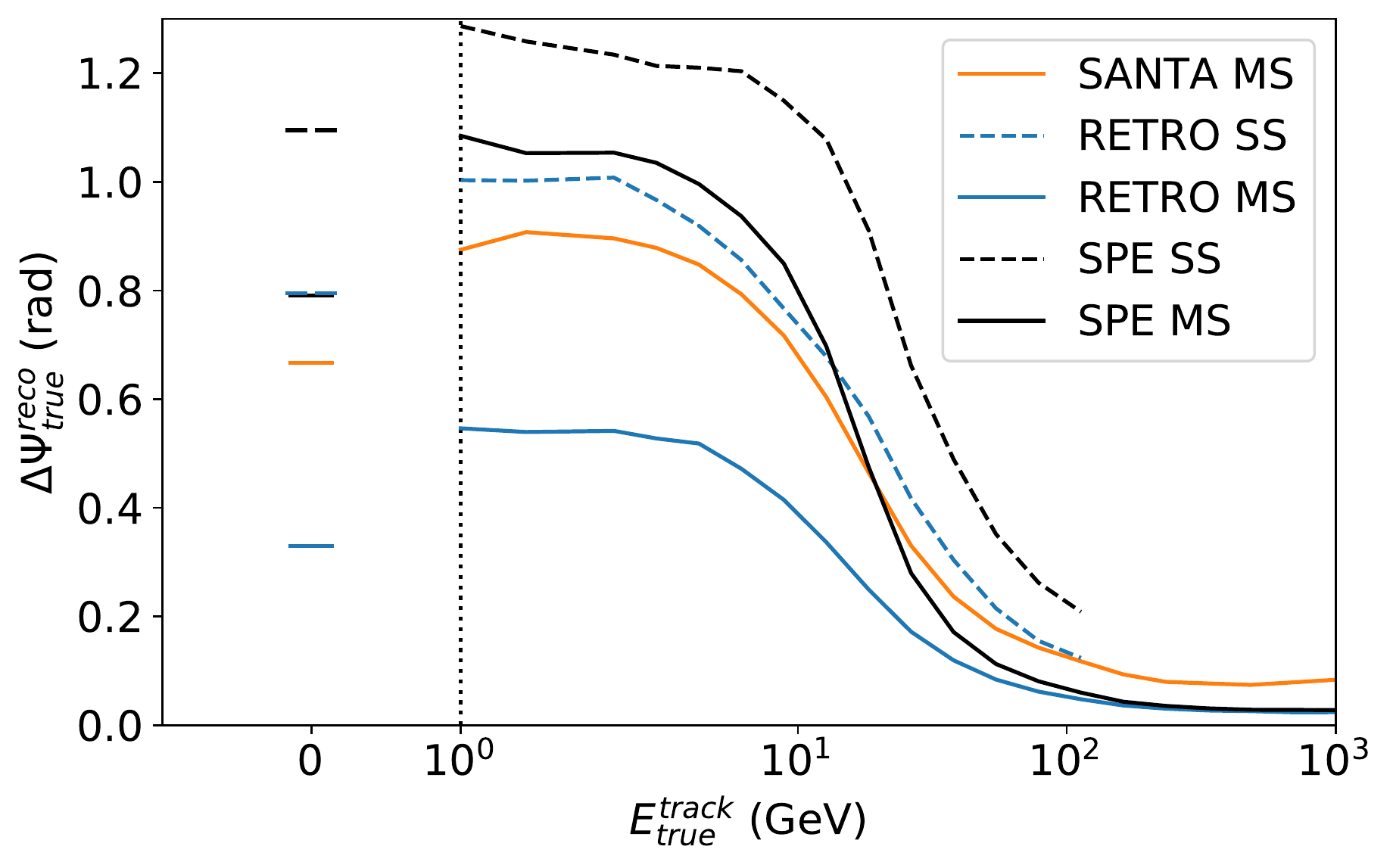}
\end{subfigure}
\caption{Angular resolutions for single-string (SS) and multi-string (MS) events. The results of \textsc{retro} (blue) and \textsc{santa} (orange) are shown together with the \textsc{spefit} (black) reconstruction. Since \textsc{santa} has no sensitivity to azimuth for single string events the respective line is not shown. The lines in the first two plots of this figure are the result of a KDE to the actual distributions. IQR is the 50\% interquantile range, i.e. the difference between the 0.25 and the 0.75 quantile. The last plot shows the median angle between true and reconstructed direction ($\Delta\Psi_{true}^{reco}$) vs. the true track energy. Single string lines are just shown up to 100\;GeV, because there are very few single string events above these energy.}
\label{fig:angular_res}
\end{figure}

Figure~\ref{fig:angular_res} shows the angular resolutions of \textsc{retro} and \textsc{santa} together with results from the simple online reconstruction \textsc{spefit}~\cite{Ahrens:2003fg}. The events are split by single string and multi-string events. The criterion if an event is classified as single or multi-string is based on \textsc{santa}, i.e. the distributions are based on identical sets of events.
For multi-string events, both reconstructions outperform \textsc{spefit}, whereas for single string events \textsc{santa} shows no sensitivity to azimuth and therefore the respective line is not included. In either of the cases, \textsc{retro} gives undoubtedly the best resolutions. The azimuth resolutions strongly benefit if more than one string sees light, since this is a measurement within the horizontal plane. For \textsc{spefit}, the interquantile range decreases by about a factor of 3 and for \textsc{retro} by nearly a factor of 4 when comparing the resolutions for single and multi-string events.
Similar to azimuth, the zenith resolution improves if multiple strings see light. For both types of events, \textsc{retro} shows again the best performance, followed by \textsc{santa}.
The last panel in Fig.~\ref{fig:angular_res} shows the median pointing error, i.e. the angle between the true and reconstructed directions, as a function of the true track energy (cascades all have track energy 0). The angular resolution increases as expected with higher track energies (longer tracks), and again a similar picture is painted with \textsc{retro} outperforming the others. The best resolutions ultimately converge to a median of around 0.03 radians (1.72°) for multi-string events with a track energy of a few hundred GeV.
The slight performance gain for zero track energy can be attributed to high energy cascades being present.

\begin{figure}[h]
    \centering
    \includegraphics[width=\linewidth]{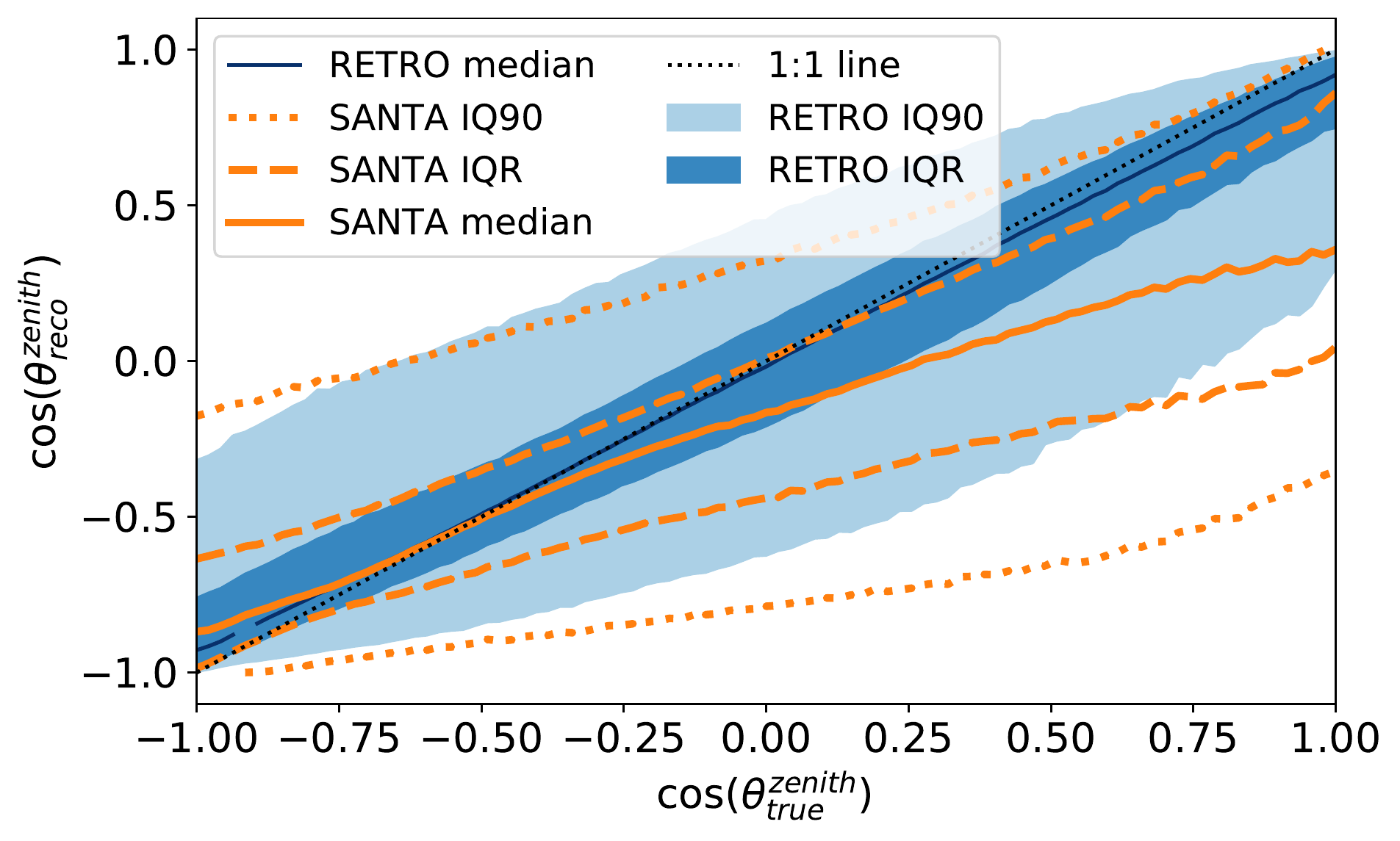}
    \caption{Reconstructed versus true cosine zenith values for \textsc{retro} (blue) and \textsc{santa} (orange). The median is shown as well as the 50 and 90 percent interquantile range (denoted as IQR and IQ90).}
    \label{fig:coszen_reco_true}
\end{figure}

Figure~\ref{fig:coszen_reco_true} shows that \textsc{santa} exhibits a bias towards up-going angles, while \textsc{retro} has virtually no bias and stays much closer to the ideal 1:1 line. The bends in the figure towards the poles result from the fact that values of the cosine are physically bounded between --1 and 1.

\subsubsection{Reconstruction time}
\begin{figure}[h]
\centering
\begin{subfigure}{.49\textwidth}
  \centering
  \includegraphics[width=\linewidth]{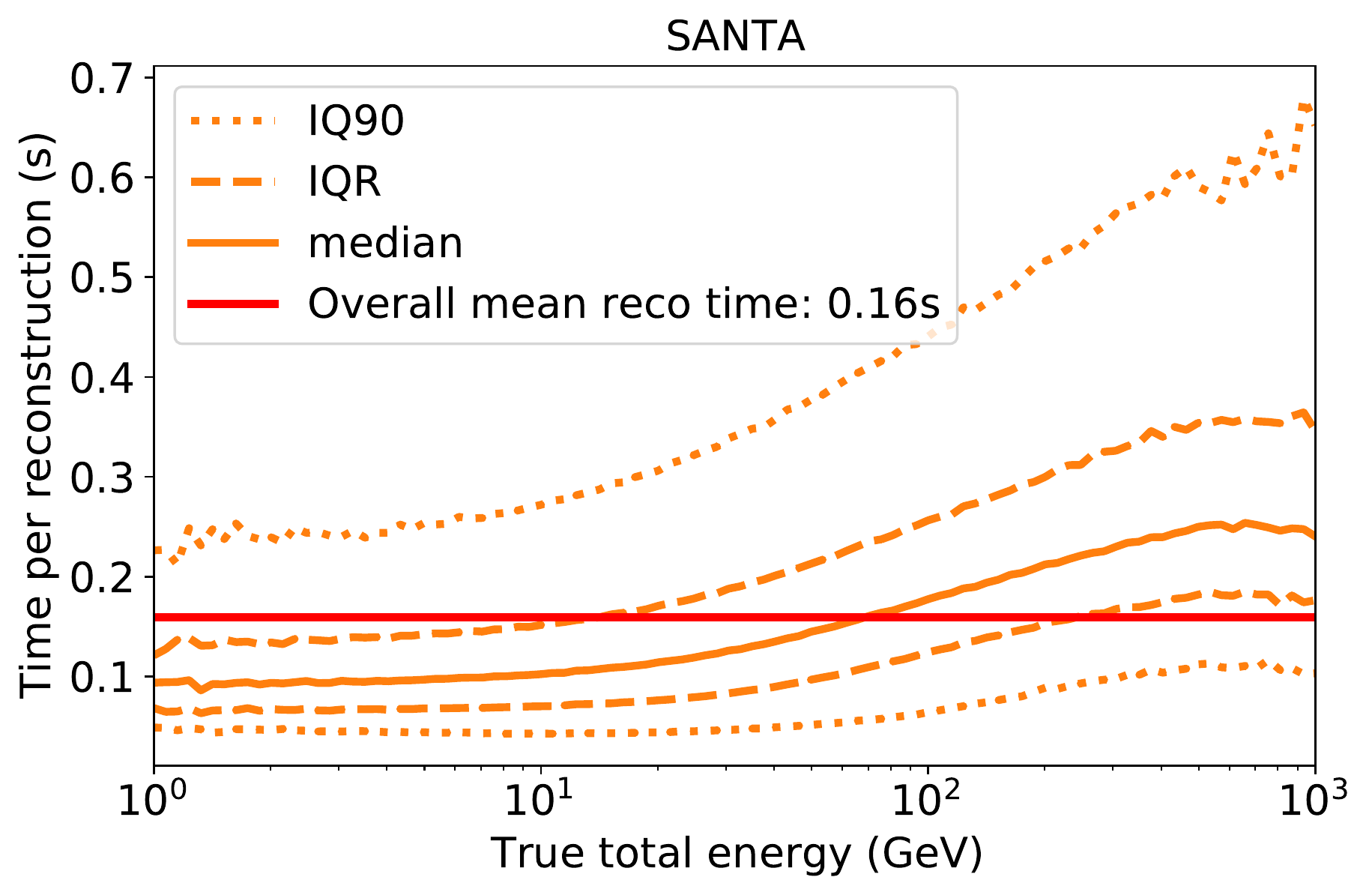}
  \caption{CPU time to reconstruct an event with \textsc{santa}. Plotted is the required CPU time, irrespective of whether the event is single string or multiple string.}
  \label{fig:santa_time}
\end{subfigure}%
\hspace{0.015\textwidth}
\begin{subfigure}{.49\textwidth}
  \centering
  \includegraphics[width=\linewidth]{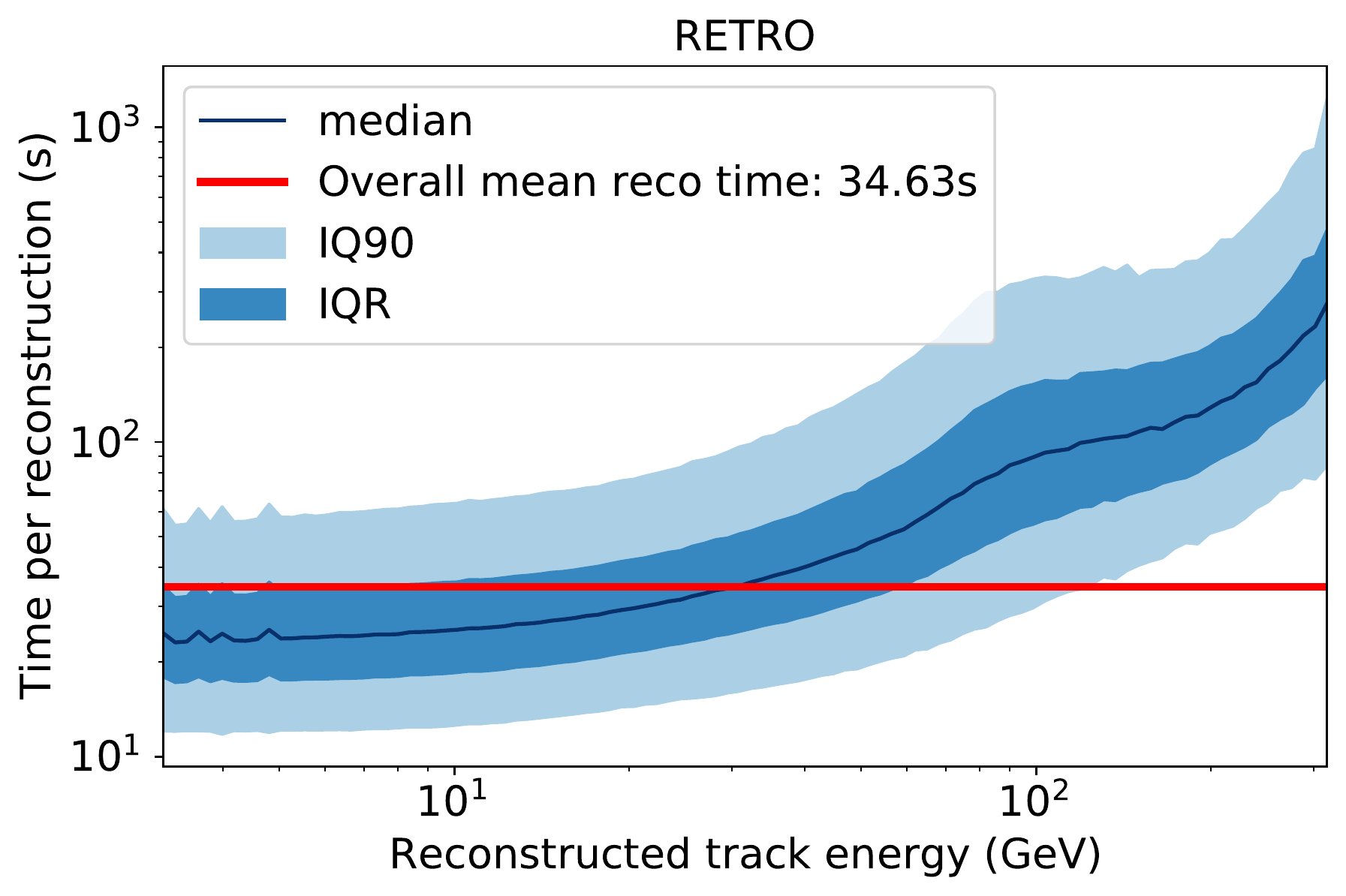}
  \caption{CPU time to reconstruct an event with \textsc{retro}. The reconstructed track energy is chosen on the x-axis because the number of calls to the \textsc{retro} photon tables mainly depends on it.}
  \label{fig:retro_time}
\end{subfigure}
\caption{Reconstructions CPU time to estimate the event parameters. The median together with the 50 and 90 percent interquantile range (IQR and IQ90) are plotted for different energies. The overall mean time is also shown as an indication of how long it takes to reconstruct a certain number of events for the event distribution used here.}
\label{fig:reco_time}
\end{figure}

Low energy MC samples in IceCube can amount up to $\mathcal{O}(10^{8}$) events, which all need to be reconstructed. Therefore, the time needed to reconstruct an event is another important property of the reconstruction performance. Figure~\ref{fig:reco_time} shows the CPU time spent per event relative to the true (total) deposited energy for \textsc{santa} and relative to the reconstructed track energy for \textsc{retro}\footnote{The reason for choosing the reconstructed track energy instead of the true total energy for \textsc{retro} is that the number of calls to the \textsc{retro} photon tables depends on the reconstructed track energy (cf. Eq.~\ref{eq:retro_llh}).}.
In this example, \textsc{santa} is around 200 times faster than \textsc{retro}, and it would typically take around 5 kCPUh to reconstruct $10^8$ events, while around 1000 kCPUh are required to do the same with \textsc{retro}.

\subsection{Full \textsc{retro} performance}
\begin{figure*}[ht]
    \centering
    \includegraphics[width=\textwidth]{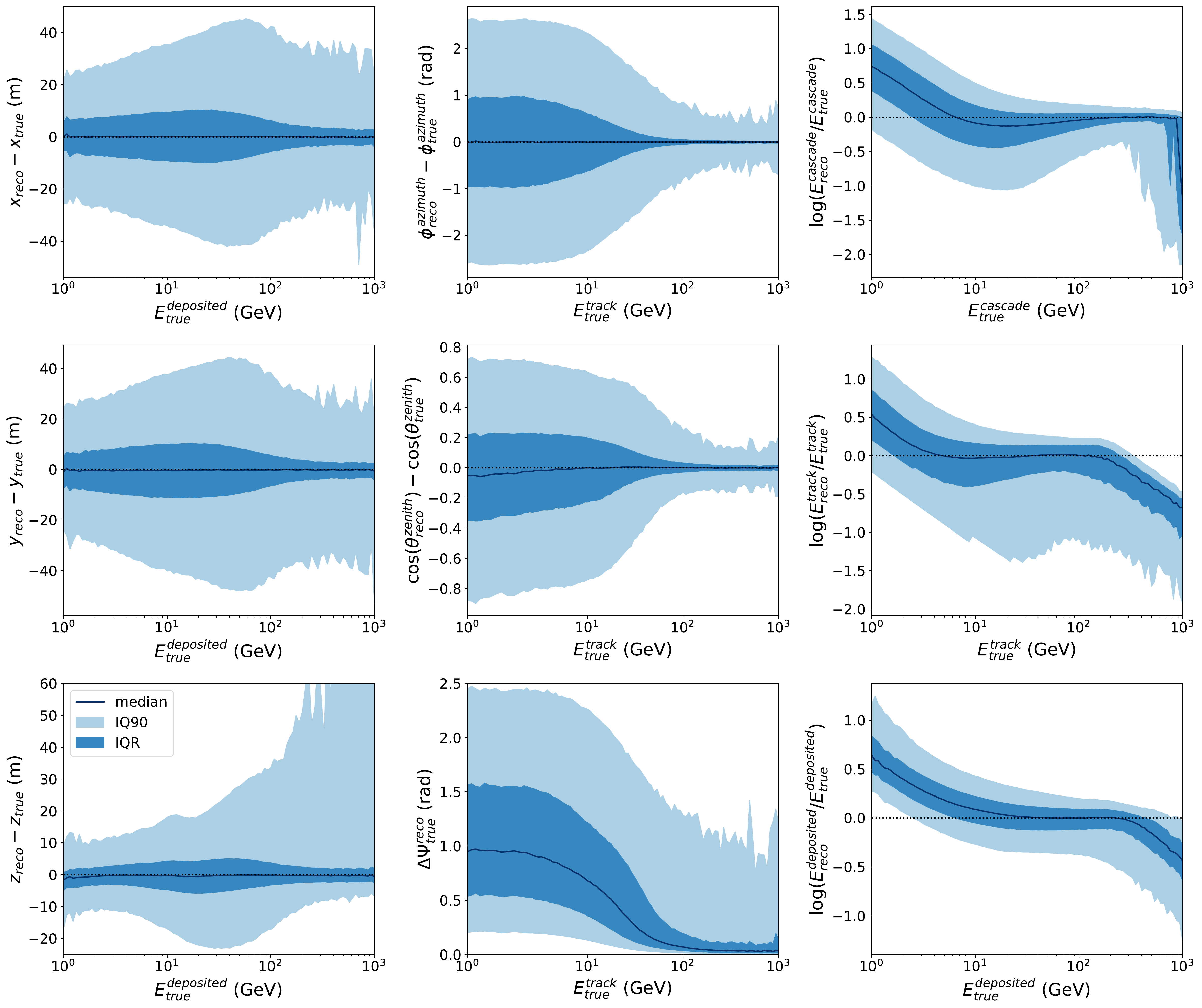}
    \caption{\textsc{retro} resolutions vs. true (total) deposited energy for vertex position and total energy. For the angular resolutions, the true track energy is used instead of the true deposited energy. In addition to azimuth and zenith resolutions also the angle between reconstructed and true direction ($\Delta\Psi^{reco}_{true}$) is shown. The single cascade and track energy resolutions are also shown, but in contrast to the other parameters vs. their respective truth. The median together with the 50 and 90 percent interquantile range (IQR and IQ90) are used to quantify the resolutions. A containment cut is used to remove all events that are reconstructed outside of DeepCore.}
    \label{fig:full_retro}
\end{figure*}

In contrast to \textsc{santa}, \textsc{retro} estimates all eight parameters used to define a neutrino interaction in IceCube DeepCore. Figure~\ref{fig:full_retro} shows the \textsc{retro} resolutions for all parameters. The resolutions of the vertex, the time and the total energy are compared versus the true (total) deposited energy. The angular resolutions are compared versus the true track energy and for track and cascade energy their respective truth is used on the $x$-axis. We use a containment cut\footnote{$-500<z_{reco}<-200$ and $\rho_{36, reco}<300$, where $\rho_{36}$ is the horizontal distance to string 36.} to remove all events that are reconstructed outside the DeepCore instrumented detector volume.

For our test sample the $x$ and $y$ vertex coordinates can be resolved at a precision of about 15\,m, while the resolution in $z$ is about 8\,m. The deteriorating resolution at high energies can be explained by partially contained events, which are high energy events outside of DeepCore that appear to be lower energy events inside or at the edge of the detector.
The angular resolutions are virtually unbiased and improve with higher track energy.
The relative energy resolutions are almost constant over the energy range shown. The bias in the energy resolution bands at very low (< 10\,GeV) and very high (> 100\,GeV) energies can be attributed to the event selection. The event sample used here is optimized for atmospheric neutrino oscillation physics with a targeted energy between 5 and 300\;GeV, meaning that the selection process is designed to pass events that appear to be in this range. Light propagation in IceCube is a stochastic process. Therefore, in some low energy events enough DOMs see light to pass the criteria. On the other side, some high energy events deposit less light than expected and also appear in the sample. These, respectively, over- and under-fluctuated events cause the bending of the energy resolution curves in the last row of Fig.~\ref{fig:full_retro}.

\section{Conclusions \& Outlook}

This article describes the state of the art in low energy reconstruction techniques as used in IceCube DeepCore for neutrino events in the GeV energy range. The algorithms must offer a high enough computational speed to be applied to up to $\mathcal{O}(10^8)$ events, and must be able to deal with very sparse data since in a typical GeV scale event around 99.7\% of all IceCube DOMs see no light.

The \textsc{santa} algorithm offers a better zenith angle reconstruction (and for multi-string events also a better azimuth reconstruction) than previous methods for events that pass a selection of unscattered photon hits. The reconstruction is in comparison fast, allowing for short turn-around times in the development of analyses and is currently used for verification purposes.

The \textsc{retro} algorithm is a full likelihood reconstruction with an eight dimensional event model and detector response. It offers superior resolutions in all reconstructed dimensions, and is applicable to any event. This significant gain in reconstruction accuracy, however, comes at the price of larger computational load---about two orders of magnitude slower than \textsc{santa}. \textsc{retro} is serving as the baseline reconstruction for current low energy oscillation, and beyond the standard model analyses such as light sterile neutrinos, non-standard interactions or dark matter searches in IceCube DeepCore.

At the same time, reconstruction algorithms remain an actively studied topic in IceCube. Especially novel techniques from machine learning are being tested as fast approximations of the detector response, surrogate likelihoods, or applied directly in the context of regression of reconstruction quantities and classification (see for example Refs.~\cite{IceCube:2021umt,IceCube:2021gpn,Minh:2021opc,Yu:2021rgp}).

\appendix

\twocolumn{
\begin{figure*}[ht]
    \centering
    \includegraphics[width=\textwidth]{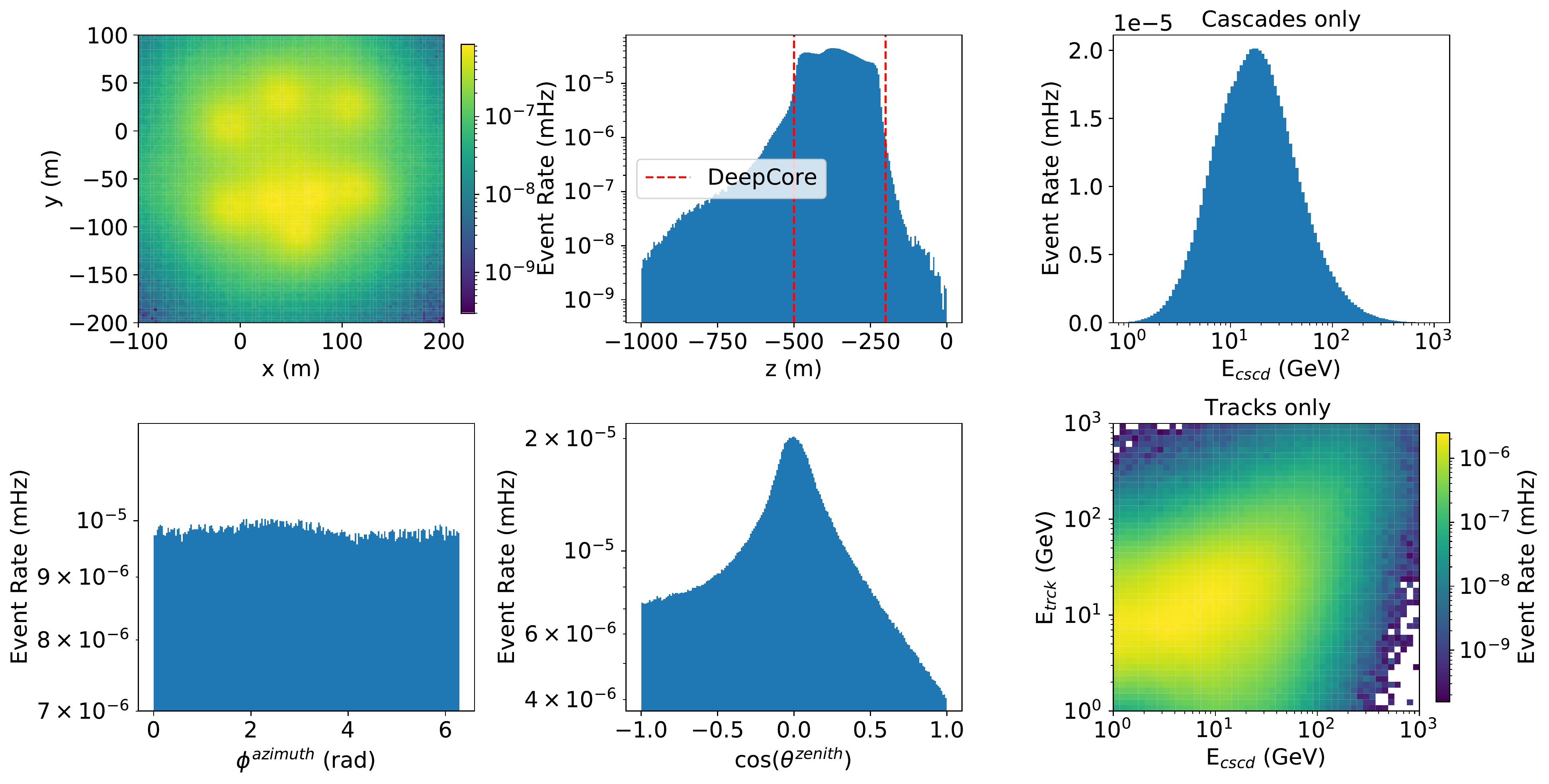}
    \caption{True parameter distributions of the full test event sample after the selection process weighted by atmospheric flux and oscillation probability. The position and angular distributions contain all events, the energy distributions are split up in a cascades only distribution (upper right) showing only the cascade energy and a track only distribution (lower right) showing cascade and track energy.}
    \label{fig:true_dists}
\end{figure*}
}

\section{Spherical CRS Optimizer}
\label{sec:crs}

The controlled random search with local mutation (\textsc{crs2}) algorithm described in \cite{Kaelo2006} is a derivative free, global optimization strategy.
We extend the existing \textsc{crs2} algorithm into our custom variant\footnote{\url{https://pypi.org/project/spherical-opt}} that is suited to correctly treat angles (azimuth and zenith).

The \textsc{crs2} algorithm, which is a simplex based optimization adapted for multi-modal target functions, uses internally geometric point reflections and centroid calculations. In the standard implementation those operations are performed assuming Euclidean geometry, however this is not the case for angles. We therefore execute these calculations for the neutrino’s zenith and azimuth values on the sphere, resulting in a different (correct) behavior. The adapted algorithm is independent of the choice of the coordinates and has better convergence than the original implementation.
The customization lies in the calculation of centroids and point reflection for the dimensions on the sphere:

\paragraph{Centroid Calculation}
\label{sec:centroid}
In order to calculate the centroid of $n$ points given their spherical coordinates $(\vartheta_1, \varphi_1)$, $(\vartheta_2, \varphi_2)$, $\dots (\vartheta_n, \varphi_n)$, we resort to their representation in Cartesian coordinates and calculate the centroid (element-wise average) in $x$, $y$ and $z$, normalize to assure $x^2 + y^2 + z^2 = 1$, and subsequently transform back to spherical coordinates to obtain the centroid in zenith and azimuth.

Also for the operations in the \textsc{crs2} algorithm where midpoints of two points have to be determined, we use the same procedure, i.e. the centroid of two points.

\paragraph{Point Reflection}
\label{sec:reflection}
Calculating geometric point reflections, as used in \textsc{crs2}, is slightly more involved. To reflect a point $p$ around the centroid $(\vartheta_c, \varphi_c)$ into a new point $\overline{p}$, we first transform the coordinates so that the centroid comes to lie at the North Pole of the unit sphere. This is achieved by a rotation $R_z^T(\phi_c)$ about the $z$-axis such that the centroid point will be on the $x-z$ plane. Then we rotate around the $y$ axis by a rotation $R_y^T(\vartheta_c)$, which brings the centroid to the desired location. In these coordinates, reflections about the centroid (now at the North Pole), are simply taking the negative values of the $x$ and $y$ dimensions of a point in its Cartesian representation. After this, we transform back into the original representation by inverse rotation in the correct order. The full operation is shown in Eq.~\ref{eq:reflection} for a point $p$ in its Cartesian representation.

\begin{widetext}

\begin{align}
\begin{split}
\boldsymbol{\overline{p}} &= R_z(\varphi_c) \cdot R_y(\vartheta_c) \cdot \mathrm{diag}(-1,-1,1) \cdot R_y^T(\vartheta_c) \cdot R_z^T(\varphi_c) \cdot \boldsymbol{p}\\
&= 
\begin{pmatrix} 
c_\varphi (-c_\varphi c_\vartheta^2 + c_\varphi s_\vartheta^2) - s_\varphi^2 & c_\varphi s_\varphi + s_\varphi (-c_\varphi c_\vartheta^2 + c_\varphi s_\vartheta^2) &   2 c_\varphi c_\vartheta s_\vartheta \\
c_\varphi s_\varphi + c_\varphi (-s_\varphi c_\vartheta^2 + s_\varphi s_\vartheta^2) & -c_\varphi^2 + s_\varphi (-s_\varphi c_\vartheta^2  + s_\varphi s_\vartheta^2) &    2 s_\varphi c_\vartheta s_\vartheta \\
2 c_\varphi c_\vartheta s_\vartheta & 2 s_\varphi c_\vartheta s_\vartheta & c_\vartheta^2 - s_\vartheta^2    \\      
\end{pmatrix}
\cdot \boldsymbol{p}, \textrm{where} 
\begin{cases}
c_\varphi = \cos{\varphi_c}\\
s_\varphi = \sin{\varphi_c}\\
c_\vartheta = \cos{\vartheta_c}\\
s_\vartheta = \sin{\vartheta_c}
\end{cases}
\label{eq:reflection}
\end{split}
\end{align}
\end{widetext}

\bibliographystyle{spphys2}       
\bibliography{references}

\begin{acknowledgements}
USA {\textendash} U.S. National Science Foundation-Office of Polar Programs,
U.S. National Science Foundation-Physics Division,
U.S. National Science Foundation-EPSCoR,
Wisconsin Alumni Research Foundation,
Center for High Throughput Computing (CHTC) at the University of Wisconsin{\textendash}Madison,
Open Science Grid (OSG),
Extreme Science and Engineering Discovery Environment (XSEDE),
Frontera computing project at the Texas Advanced Computing Center,
U.S. Department of Energy-National Energy Research Scientific Computing Center,
Particle astrophysics research computing center at the University of Maryland,
Institute for Cyber-Enabled Research at Michigan State University,
and Astroparticle physics computational facility at Marquette University;
Belgium {\textendash} Funds for Scientific Research (FRS-FNRS and FWO),
FWO Odysseus and Big Science programmes,
and Belgian Federal Science Policy Office (Belspo);
Germany {\textendash} Bundesministerium f{\"u}r Bildung und Forschung (BMBF),
Deutsche Forschungsgemeinschaft (DFG),
Helmholtz Alliance for Astroparticle Physics (HAP),
Initiative and Networking Fund of the Helmholtz Association,
Deutsches Elektronen Synchrotron (DESY),
and High Performance Computing cluster of the RWTH Aachen;
Sweden {\textendash} Swedish Research Council,
Swedish Polar Research Secretariat,
Swedish National Infrastructure for Computing (SNIC),
and Knut and Alice Wallenberg Foundation;
Australia {\textendash} Australian Research Council;
Canada {\textendash} Natural Sciences and Engineering Research Council of Canada,
Calcul Qu{\'e}bec, Compute Ontario, Canada Foundation for Innovation, WestGrid, and Compute Canada;
Denmark {\textendash} Villum Fonden and Carlsberg Foundation;
New Zealand {\textendash} Marsden Fund;
Japan {\textendash} Japan Society for Promotion of Science (JSPS)
and Institute for Global Prominent Research (IGPR) of Chiba University;
Korea {\textendash} National Research Foundation of Korea (NRF);
Switzerland {\textendash} Swiss National Science Foundation (SNSF);
United Kingdom {\textendash} Department of Physics, University of Oxford.
\end{acknowledgements}

\end{document}